\documentclass[12pt]{elsarticle}
\usepackage{graphicx}
\graphicspath{{./images/}}
\usepackage{url}
\usepackage{multirow}
\usepackage[table,xcdraw]{xcolor}
\usepackage{tabularx}
\usepackage{amsmath}
\usepackage{caption}
\usepackage{subcaption}
\usepackage{hhline}
\usepackage{verbatim}

\biboptions{sort&compress}

\usepackage{amssymb}

\usepackage{xcolor}
\usepackage{soul}

\begin{document}

\begin{frontmatter}



\title{Phonological Level wav2vec2-based Mispronunciation Detection and Diagnosis Method}


\affiliation[inst1]{organization={School of Electrical and Computer Engineering},
            addressline={University of New South Wales}, 
            city={Sydney},
            postcode={2052}, 
            state={NSW},
            country={Australia}}

\author[inst1]{Mostafa Shahin}
\author[inst1]{Julien Epps}
\author[inst1]{Beena Ahmed}

\begin{abstract}
The automatic identification and analysis of pronunciation errors, known as Mispronunciation Detection and Diagnosis (MDD) plays a crucial role in Computer Aided Pronunciation Learning (CAPL) tools such as Second-Language (L2) learning or speech therapy applications. Existing MDD methods relying on analysing phonemes can only detect categorical errors of phonemes that have an adequate amount of training data to be modelled. With the unpredictable nature of the pronunciation errors of non-native or disordered speakers and the scarcity of training datasets, it is unfeasible to model all types of mispronunciations. Moreover, phoneme-level MDD approaches have a limited ability to provide detailed diagnostic information about the error made. In this paper, we propose a low-level MDD approach based on the detection of speech attribute features. Speech attribute features break down phoneme production into elementary components that are directly related to the articulatory system leading to more formative feedback to the learner.
We further propose a multi-label variant of the Connectionist Temporal Classification (CTC) approach to jointly model the non-mutually exclusive speech attributes using a single model. The pre-trained wav2vec2 model was employed as a core model for the speech attribute detector.
The proposed method was applied to L2 speech corpora collected from English learners from different native languages. The proposed speech attribute MDD method was further compared to the traditional phoneme-level MDD and achieved a significantly lower False Acceptance Rate (FAR), False Rejection Rate (FRR), and Diagnostic Error Rate (DER) over all speech attributes compared to the phoneme-level equivalent.
\end{abstract}


\end{frontmatter}


\section{Introduction}
\label{sec:intro}
Unlike acoustic modeling for Automatic Speech Recognition (ASR) applications where an abstract model that can handle all variations of the same word is desirable, pronunciation assessment applications such as second language (L2) learning, speech therapy, and language proficiency tests, require accurate detection of any deviation from standard pronunciation.

Mispronunciation detection and diagnosis (MDD) aims to automatically detect pronunciation errors in speech productions and diagnose them by providing error details such as the error location and type as well as the pronunciation quality score.  MDD is a key feature of Computer-Aided Pronunciation Learning (CAPL) systems.

The level of diagnostic information provided by the MDD model depends on the level of evaluation the trained model can perform which in turn is restricted by the level of annotation of the training data. Only if the annotation is at the phoneme-level, i.e., speech is annotated with its pronounced phoneme sequence, can a phoneme-level model be trained and used to provide phoneme-level diagnostic information, e.g., the location of the mispronounced phoneme and the type of the error (substitution, deletion, or insertion).

Phoneme-level assessment is the most common assessment level used in MDD \cite{RN30,RN31,RN37,RN41,RN42,RN43,RN44,RN45}. Different approaches have been proposed to achieve phoneme-level MDD, including the scoring approach \cite{RN43, RN112, RN113, RN114, RN115, RN116, RN117, RN118}, the rule-based approach \cite{RN33, RN34, RN42, RN37}, the classification approach \cite{RN120, RN121, RN122, RN123, RN124, RN125, RN126, RN127, RN46}, and the free-phoneme recognition approach \cite{RN32, RN95, RN132, RN133}.

One of the limitations of phoneme-level MDD is that it can only diagnose categorical pronunciation errors or mispronounced phonemes that exist in the acoustic model. Uncategorical errors, where the pronounced phoneme is a distorted version of the modelled phoneme or a new phoneme borrowed from other languages, are difficult to detect with phoneme-level MDD. To handle uncategorical errors, the acoustic model needs to include phonemes from multiple languages as well as all possible pronunciation variations of each phoneme, which is infeasible. Moreover, diagnostic information obtained from the phoneme-level MDD can't be used to construct formative feedback with corrective instruction to the learner.

In this work, we propose a low-level MDD system that detects and diagnoses pronunciation errors at the speech attribute level. Speech attributes, such as manners and places of articulations, provide a low-level description of sound production in terms of which articulators are involved and how these articulators move to produce a specific sound. Any alteration in these attributes causes a pronunciation error. Therefore, accurate modeling of these attributes instead of phonemes can pave the way to a fully automated and interactive pronunciation assessment application where the learner receives informative and diagnostic automatic feedback not only about the existence of incorrect pronunciation, but also how the error is made. Furthermore, modeling speech attributes can be performed solely using typically pronounced datasets which are abundantly available, unlike atypical datasets such as disordered or non-native speech. Additionally, speech attributes are common across most spoken languages enabling modeling with speech corpora from multiple languages \cite{RN146}. 

Our proposed speech attribute detection model is a wav2vec2-based Sequence-to-Sequence (Seq2Seq) classification model trained using a multi-label variant of the Connectionist Temporal Classification (CTC) criteria to handle the non-mutually exclusive nature of speech attributes, i.e, the same phoneme can be characterized by multiple speech attributes. To validate our model, it was applied to English as L2 speech corpora collected from different L1 speakers. Additionally, we compared the detection and diagnostic accuracy of a state-of-the-art phoneme-level MDD method with our proposed speech attribute-based approach. We also conducted experiments to demonstrate the low-level diagnosis capability of the model and the potential formative feedback message it can provide.

This paper presents three key contributions. Firstly, we established a new benchmark for speech attribute detection by utilizing the wav2vec2 speech representation upstream model. Secondly, we introduced a novel multi-label CTC approach to simultaneously learn 35 non-mutually exclusive speech attributes. Lastly, we proposed a low-level MDD method based on our speech attribute detection model.

The paper is structured as follows: In Section 2, the limitations of current phoneme-level MDD approaches are discussed. Section 3 presents our proposed method, followed by a description of the speech corpora used in Section 4. The experimental setup is outlined in Section 5, while the results are demonstrated and discussed in Section 6. Finally, Section 7 concludes the paper.

\section{Related Work}
\subsection{Mispronunciation Detection and Diagnosis (MDD)}
The Goodness Of Pronunciation (GOP) algorithm \cite{RN43} is the earliest and most successful phoneme-level speech assessment method utilized in several applications to measure phoneme-level pronunciation quality \cite{RN10, RN65, RN88, RN108, RN109, RN110, RN111}. The GOP approximates the posterior probability of each phoneme by taking the ratio between the forced alignment likelihood and the maximum likelihood of the free-phone loop decoding using a Gaussian Mixture Model-Hidden-Markov Model (GMM-HMM) acoustic model. This score is then used as a confidence score representing how close the pronunciation is to the target phoneme. Subsequent methods have leveraged DNN acoustic modelling and proposed a DNN-based method to estimate the GOP \cite{RN117, RN118, RN119}.

Despite its success, the GOP algorithm is very sensitive to the quality of the acoustic model used. The acoustic model affects not only the estimate of the posterior probability but also the accuracy of time boundaries obtained from forced alignment. In addition, as the decision threshold is determined using a mispronunciation dataset, it can be error specific and thus very hard to generalise to different types of pronunciation errors.

Phoneme-level error detection has also been treated as a binary classification problem, with each phoneme classified as “correct” or “mispronounced” using conventional classification methods such as SVM \cite{RN122}, decision tree \cite{RN120}, Linear Discriminant Analysis (LDA) \cite{RN121}, DNN \cite{RN126}, etc. Mel Frequency Cepstral Coefficients (MFCCs) are the most common features used with the classification methods \cite{RN120}. Formant frequencies and GOP scores were also utilised to classify between correct and incorrect phoneme pronunciation \cite{RN123}. 

Although all these classifiers led to a significant improvement compared with confidence score methods such as GOP, they still need large amounts of accurately annotated non-native data to model the mispronounced phonemes. Moreover, the mispronounced data needs to include all possible pronunciation errors, which is usually not feasible to collect.

In \cite{RN44}, an anomaly detection-based model was trained solely on the standard English pronunciation, namely native English speakers, and tested on foreign-accented speech and disordered speech. The method treated the mispronunciation as a deviation (anomaly) from the standard pronunciation. To detect the anomalies a One-Class SVM (OCSVM) model was trained for each phoneme using speech attribute features, namely manners and places of articulation. The method was shown to outperform the DNN-based GOP method in both disordered and foreign-accented speech.
Recently, \cite{RN46} investigated fine-tuning a self-supervised speech representation pre-trained model, namely wav2vec2 \cite{RN127} to perform phoneme level binary classification of L2 speech pronunciation as correct/mispronounced.
Both the scoring and classification approaches can only provide either a soft (score) or hard (binary) decision on the phoneme pronunciation without any detailed description of the type of error. 

The Extended Recognition Network (ERN) method was proposed to provide more detailed descriptions of the pronunciation error. It can identify the location of the mispronounced phoneme, and the type of the error, and recognise the erroneous phoneme. Unlike forced alignment used in the scoring method which contains only the canonical phonetic transcription of the prompt word, the ERN extends the single path alignment to a network that contains the correct (canonical) path and the expected mispronounced paths on phoneme level based on phonological rules designed for a specific learning domain \cite{RN33, RN37, RN42}.

The design of the search network is crucial for the reliability of the ERN-based MDD systems. Hand-crafted phonological rules are the most common designing criterion for the ERN \cite{RN42, RN37, RN45, RN33, RN34}. Data-driven phonological approaches have also been proposed to automatically create the ERN. \cite{RN94} first performed an automatic phonetic alignment step between the canonical phonetic transcription and the corresponding L2 transcription then all pairs of mismatched phonemes along with their contextual phones were grouped to form the initial set of rules. Finally, a rule selection criterion was performed to select the most significant rules. In \cite{RN129}, the authors proposed a grapheme-to-phoneme-like model trained on phonetic transcriptions of L2 learners.

The decoding of the ERN is performed by an acoustic model that is trained either on standard pronunciation speech corpora, such as L1 native speakers \cite{RN42, RN94}, or non-standard pronunciation speech corpora, such as L2 non-native speakers \cite{RN130, RN131}, or child disordered speech \cite{RN37}. As the ERN method was proposed around 20 years ago, the most common acoustic model used was the GMM-HMM acoustic model \cite{RN33, RN42, RN94, RN128, RN130}. Later the DNN-HMM acoustic model was utilised \cite{RN34, RN37, RN45, RN131} and shown to outperform the GMM-HMM in the ERN methods specifically when trained on a small non-standard dataset \cite{RN37}. Moreover, the standard acoustic model has been adapted to non-standard domains using techniques such as Maximum Likelihood Linear Regression (MLLR) for GMM-HMM-based models \cite{RN33, RN128} or transfer learning for DNN-HMM-based acoustic models \cite{RN45}. \cite{RN130} proposed a discriminative training criterion to directly optimise the acoustic model for MDD. The authors incorporated False Acceptance Rate ($FAR$), False Rejection Rate ($FRR$), and Diagnostic Error Rate ($DER$) in the acoustic model objective function and trained it using L2 non-native speech corpora.

With significant improvements in End-to-End (End2End) deep learning-based acoustic models, most of the current SOTA MDD approaches have adopted free-phone recognition criteria. Unlike the ERN, free-phone recognition has no restricted search space and therefore any pronounced phoneme sequence can be captured. That is, free-phone recognition MDD systems work by estimating the pronounced phoneme sequence and the evaluation of the system is achieved by aligning the recognised phoneme sequence with the annotated one. The free-phoneme recognition model is adapted to the pronunciation learning domain by incorporating linguistic information from the prompts used in pronunciation learning which, in most cases, are pre-designed and available. This linguistic information is extracted on the character level \cite{RN31}, phonemic level \cite{RN32, RN95, RN132, RN133}, graphemic level \cite{RN95}, or a combination of different levels \cite{RN95}.

In \cite{RN30} the CNN-RNN-CTC system was proposed as an early attempt of an End2End MDD method. The system has a straightforward phoneme recognition architecture trained using a combination of native and L2 speech corpora and makes use of the Connectionist Temporal Classification (CTC) loss to avoid direct alignment between the phoneme sequence and the speech signal.

Recent methods adopted an encoder-decoder mechanism to achieve MDD by using multiple encoders to encode the audio signal \cite{RN136} along with the prompt sentence phonemes \cite{RN132, RN32}, characters \cite{RN31}, or words \cite{RN135}. The encoder architecture is commonly a stacked CNN-RNN \cite{RN31, RN32} while the decoder is mostly an attention-based network.

All the previous methods use hand-crafted features extracted from a short-time speech window, however, several recent SOTA speech recognition systems leverage the raw speech signal and use a learnable feature extraction network that can be integrated and trained within the whole network \cite{RN138, RN139, RN140}.

One of the shortcomings of the free-phoneme recognition method is that the phoneme sequence output is limited to the modelled phoneme set. Therefore, it is not able to detect distortion errors where the learner pronounces a distorted version of a phoneme that cannot be explicitly replaced with another phoneme within the target language phoneme set. This commonly occurs when the L2 learner is influenced by their L1 language.

In pronunciation learning, pronunciation errors made by the learner can be unpredictable. For instance, in L2 learning pronunciation errors are influenced by the proficiency of the learner and the degree of discrepancy between L1 and L2. Therefore, there are high variations in the way that the learner will adapt their pronunciation to try and match the target pronunciation. Given the limited amount of annotated mispronounced speech data available, it is infeasible to model all these variations.
Existing MDD methods can only diagnose categorical pronunciations that can be modelled by the acoustic model, and struggle otherwise. Scoring approaches such as the GOP can help in detecting mispronunciations if the deviation from the correct pronunciation is high enough, however, the actual pronounced error and its type cannot be determined.

Here we have thus proposed a low-level MDD based on the speech attribute features. Speech attribute features, also known as phonological features, can break down the phoneme into elementary components that form the phoneme. These features include manners and places of articulations that describe the articulators’ positions and movements during sound production. There are several advantages of using speech attributes in MDD: 1) the speech attributes are only limited by the possible positions of the articulators and thus shared among most spoken languages, 2) their models can be solely trained using correctly pronounced speech which alleviates the need for annotated mispronounced data, and 3) they can provide lower-level diagnostic information describing how the error is made and which attribute is missing allowing more formative feedback to be provided.

\subsection{Speech Attribute Modelling}
Speech attributes have been successfully utilized in various domains such as improving ASR performance \cite{RN288, RN289}, language identification \cite{RN146}, speaker verification \cite{RN290} and Mispronunciation Detection and Diagnosis (MDD) \cite{RN154}. However, the speech attribute detection models are commonly trained at the frame level and hence frame-level labelling of the speech signal is required for all training samples. The attributes are obtained by first performing a forced alignment on a phoneme level and mapping phonemes to their corresponding attributes. The performance of the resultant speech attribute model is thus highly dependent on the quality of the acoustic model used in the forced alignment process. To overcome this constraint, some studies have explored using the CTC learning criteria, which eliminates the need for prior alignment between the input signal and output label \cite{RN291, RN292, RN293, RN162}.

The speech attributes are non-mutually exclusive; each phoneme can appear in multiple attributes. For example, the phoneme /z/ is fricative, voiced and alveolar. Consequently, modelling all attributes with a single model becomes a multi-label classification problem. Existing approaches address this by constructing multiple models, with each model dedicated to an individual attribute \cite{RN148, RN149}, or a set of mutually exclusive attributes \cite{RN160,RN161}. Unfortunately, using multiple models is impractical specifically in real-time applications. In the inference time, multiple models need to be decoded which increases the application latency and consumes large memory making using speech attributes unfeasible for applications that need instant feedback to be provided to the user.

In this work, we proposed a multi-label CTC method to train a single model that can detect multiple non-mutually exclusive speech attributes. We also investigate using wav2vec2 speech representations to perform speech attribute detection as a downstream task. 
 
\section{Method}
\label{sec:meth}
\begin{figure}[h!]
    \centering    \includegraphics[width=\textwidth, height=0.35 \textwidth]{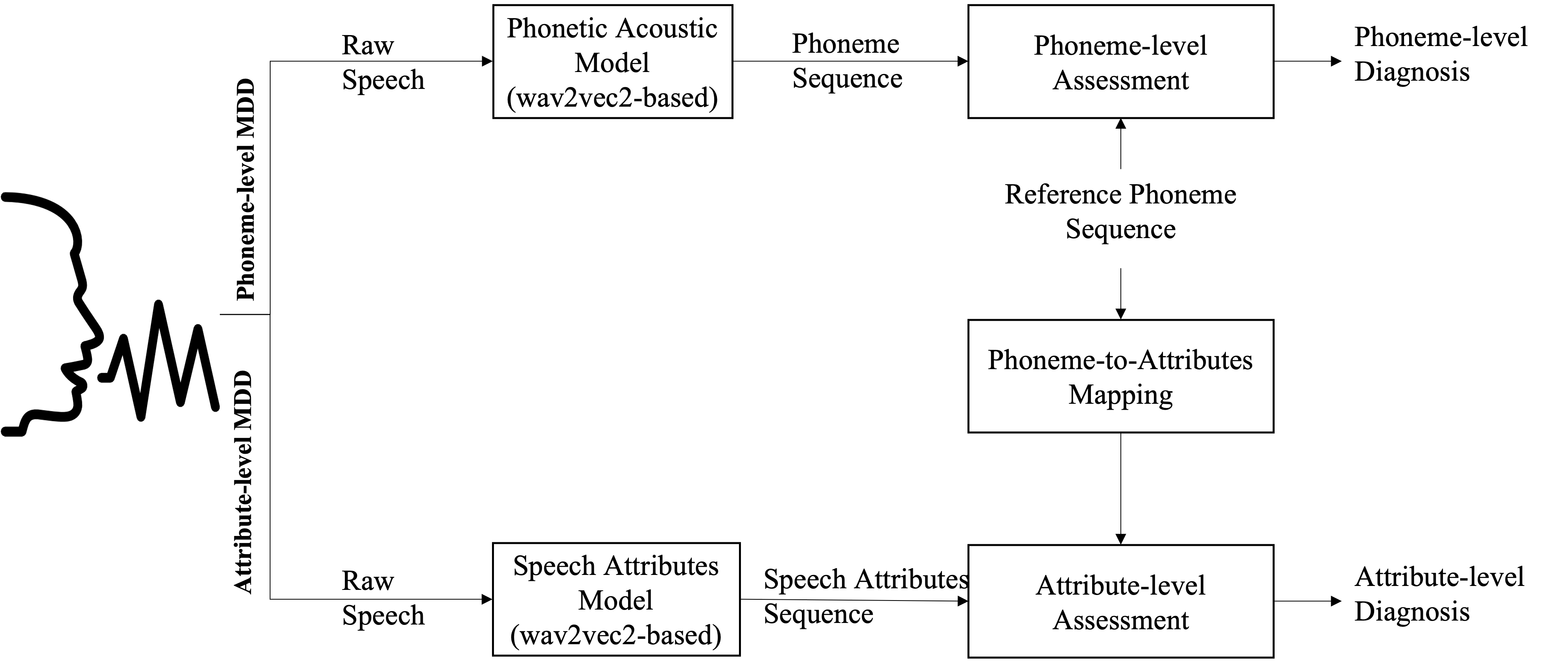}
    \caption{ A block diagram of the phoneme and attribute level Mispronunciation Detection and Diagnosis (MDD) approaches implemented. The wav2vec2-based phonetic acoustic model processes the raw speech signal and outputs a sequence of recognised phonemes. The wav2vec-based speech attributes model processes the raw speech signal and produces multiple speech attribute sequences, one for each targeted attribute. The assessment was performed at the phoneme level by aligning the recognised phoneme sequence with the reference one, and at the speech attributes level by aligning each recognised attribute sequence of $+att$/$-att$ with the corresponding reference attribute sequence. }
    \label{fig:method1}
\end{figure}

Two MDD systems were implemented, a conventional phoneme-based MDD similar to the wav2vec2-based model introduced in \cite{RN127} and the proposed speech attribute-based, with the former used as a baseline to compare the performance of our proposed MDD. A block diagram describing the two methods is depicted in Figure \ref{fig:method1}. In the phoneme-based method, the raw speech signal was processed by the phonetic acoustic model to generate a sequence of recognised phonemes. The resultant phoneme sequence was then aligned with the reference (canonical) phoneme sequence to identify and diagnose the pronounced errors at the phoneme level. In the second speech attribute-based method, the raw speech signal was processed by the speech attributes model to generate multiple sequences of $+att/-att$ representing the existence or absence of each attribute.   The reference phoneme sequence was obtained by first mapping to multiple reference speech attribute sequences of $+att/-att$. For each attribute, the reference and recognised sequences were then aligned to identify the missing, inserted, or replaced attributes at each position. 

\begin{figure}[h!]
    \centering
    \includegraphics[width=0.9 \textwidth, height=0.55 \textwidth]{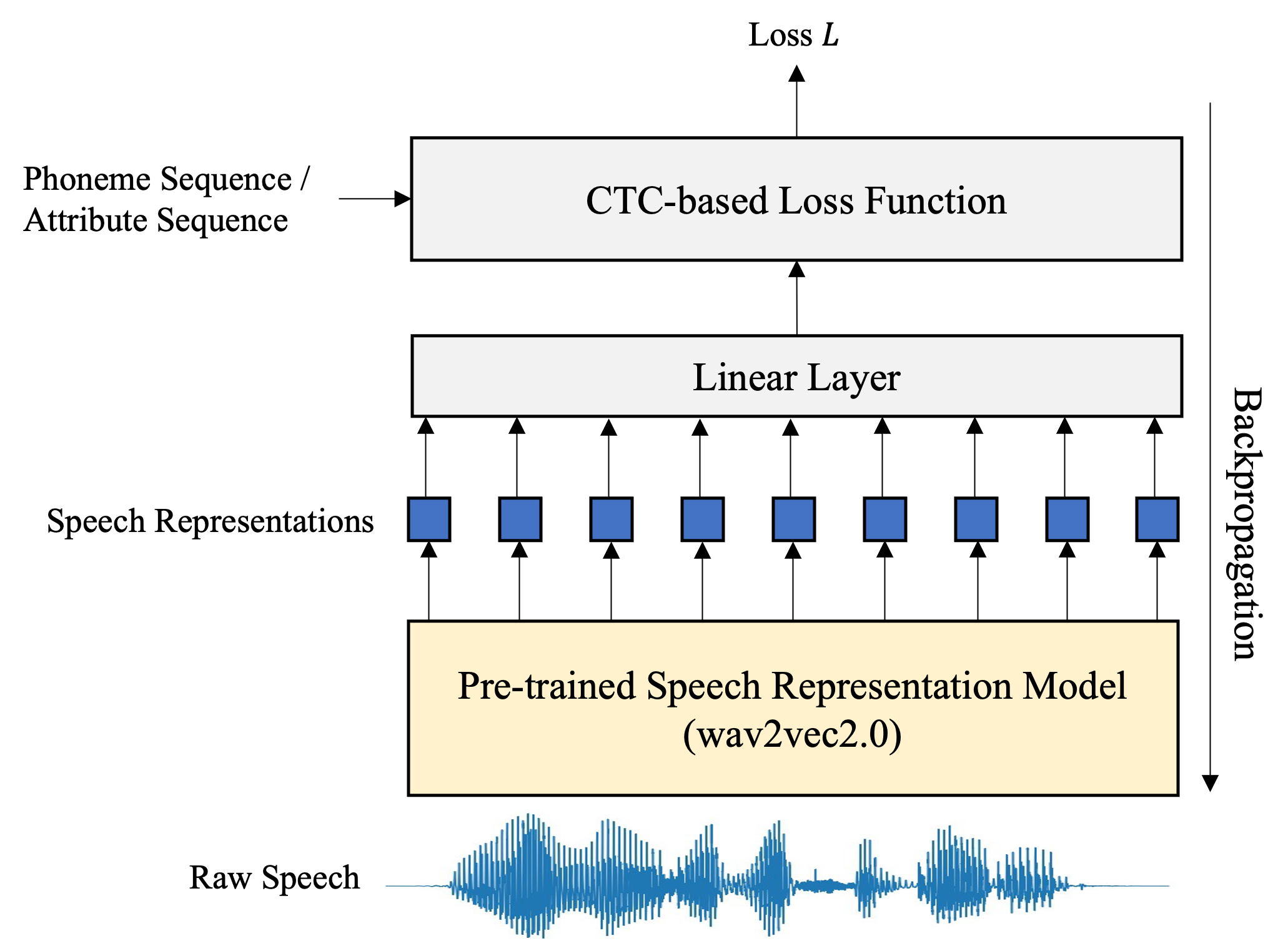}
    \caption{The architecture of the phoneme/attributes model. The speech representations generated by the pre-trained wav2vec2 model were passed to a linear layer with number of nodes equals to the number of target phonemes/attributes. The model was fed by the raw speech signal and the CTC-based loss function was calculated by considering all possible alignments between the input signal and the target phoneme/attribute sequence.
}
    \label{fig:method1-2}
\end{figure}

As shown in Figure \ref{fig:method1-2}, the wav2vec2 architecture was used as the backbone of both the speech attribute and the phonetic acoustic models while the Connectionist Temporal Classification (CTC) criterion was adopted to compute the training loss.

\begin{figure}[h!]
    \centering
    \includegraphics[width=0.9 \textwidth]{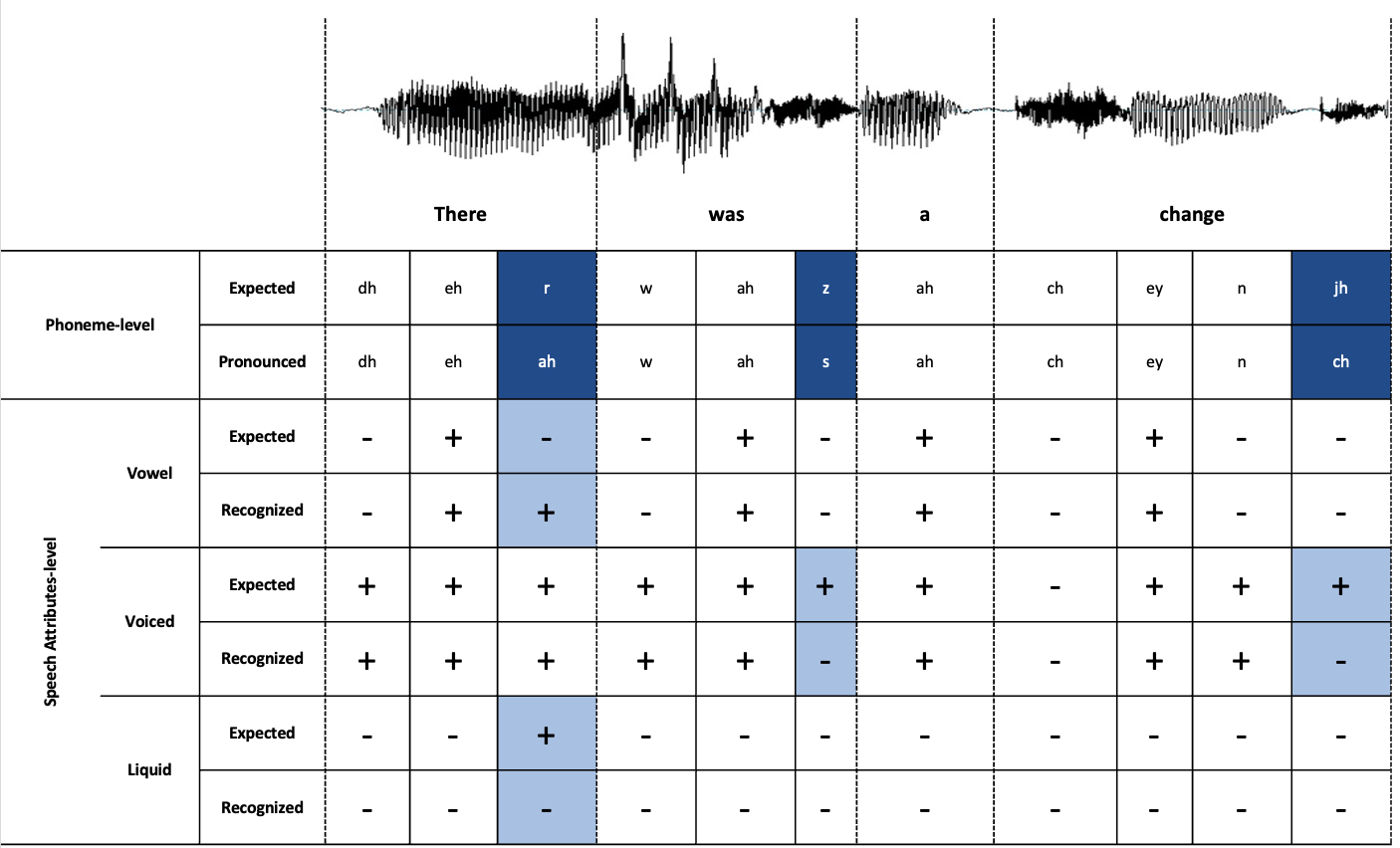}
    \caption{Mispronunciation detection and diagnosis example of sentence \emph{“There was a change”}. /r/, /z/, and /jh/ phonemes are pronounced as /ah/, /s/, and /ch/ respectively. /r/ is $-vowel$ and $+liquid$ while /ah/ is $+vowel$ and $-liquid$. Both /z/ and /jh/ are $+voiced$ while their associated erroneous phonemes /s/ and /ch/ are $-voiced$. The proposed MDD breaks down the pronunciation error into elementary components (attributes) and identifies the incorrect attribute.}
    \label{fig:method2}
\end{figure}

\clearpage
Figure \ref{fig:method2} demonstrates an example of MDD of three substitution errors detected in the production of the sentence “There was a change” by an L2 speaker. Here the consonant sound /r/ was pronounced as /ah/ vowel while the voiced /z/ and /jh/ sounds were replaced with the voiceless /s/ and /ch/.  As seen, the speech attribute detection model provides a detailed description of the pronunciation errors in terms of the manner and places of articulation.

\subsection{Connectionist Temporal Classification (CTC)}
The Connectionist Temporal Classification (CTC) loss function, initially proposed for speech recognition \cite{RN159}, enables learning of a Seq2Seq model without explicit alignment between the input and target sequences. Therefore, it has become the state-of-the-art learning criterion for several Seq2Seq models in various domains, including computer vision \cite{RN226},  handwritten text recognition \cite{RN227}, and speech recognition \cite{RN228}.
The goal of the CTC algorithm is to compute $p(l|x)$ where $x=(x^1,x^2,x^3,…x^T)$ is an input sequence of length $T$ and $l=(l^1,l^2,l^3,…l^U)$ of length $U$ its corresponding target sequence, with $U \leq T$.

$L$ defines a finite alphabet containing all possible labels where $l^t \in L$. CTC also defines a \emph{blank} label which is used when no output is assigned to a specific time slot and to differentiate time slots that belong to the same label from time slots that belong to a repeated label. Hence, let $L'=L \cup {blank}$. The output tensor $y$ of the network is therefore of size $T \times |L'|$ where $|L'|$ is the total number of possible labels in addition to the \emph{blank} output. The probability of each element $i \in L'$ at time $t$ is denoted as $y_i^t$ , with the softmax operation performed over $y^t$ to give $\sum _{i=1}^{|L'|}y_i^t=1$. 

 If $\pi$ is a label sequence of length equal to $T$ and labels $\pi^{t} \in L'$, the probability of producing an output sequence $\pi$ is computed as in (\ref{eq:CTC1})
 \begin{equation}
    \label{eq:CTC1}
     p(\pi|x)=\prod_{t=1}^T y_{\pi^t}^{t}
 \end{equation}

CTC also defines a many-to-one mapping $\beta$  that maps multiple label sequences of length $T$ and labels $\pi^t \in L'$ to one label sequence $l$ of length $U \le T$ and labels $l^t \in L$. The mapping is performed by removing \emph{blank} and repeated labels. For example, $\beta(a-ab-)=\beta(aa-ab)=\beta(-a-ab)=aab$ where $-$ denotes \emph{blank}. Therefore, the probability of a label sequence $l$ given an input sequence $x$ can be computed as the sum of all paths mapped to $l$. 

\begin{equation}
    \label{eq:CTC2}
    p(l|x)=\sum_{\pi \in \beta^{-1}(l)} p(\pi|x)
\end{equation}

\subsection{Connectionist Temporal Classification for Multi-label Sequences}
As the speech attribute features are non-mutually exclusive, where each phoneme is described by more than one attribute, the speech attribute detection becomes a multi-label classification problem. Each speech utterance can thus be mapped to different attribute sequences. The traditional CTC method can only handle single-label inputs and therefore a multi-label variant of the CTC is needed to jointly model all speech attributes.

Two approaches were introduced to handle multi-label classification problem using CTC criteria, the Separable CTC (SCTC) and the Multi-label CTC (MCTC) \cite{RN281}. The SCTC works by computing the CTC loss over each labelling category separately and then adding them together (i.e. multiplying the conditional probabilities) to get the target loss. In \cite{RN283}, the authors used the SCTC criteria to train the CTC-based model to recognise both phones and tones in multilingual and cross-lingual scenarios.

On the other hand, the MCTC first computes the probability of each target element from its categorical components and then computes the CTC loss over the target sequence \cite{RN281}. The MCTC approach has been successfully applied to polyphonic music audio to detect pitch classes \cite{RN282} and to handwritten text recognition \cite{RN281}.

Wigington et. al  \cite{RN281} discuss the two approaches for the multi-label CTC. One major issue with SCTC is that each category is treated separately and therefore it is not guaranteed that at each frame the correct combination of components exists.

In this work, we adopted the SCTC approach as the objective is the classification of the separate speech attribute categories. However, to maintain the alignment between components, one blank node was shared among all categories. The reasoning behind using a shared blank token is that in speech attribute modelling, as the categories are binary representations of each attribute, each phoneme has one and only one representation in each category. The blank token over all categories thus represents the silence frames or the transition from one phoneme to another. Using a shared blank, increases the likelihood that all categories will produce a blank at the same time frame. We refer to this approach as Separable CTC with Shared Blank (SCTC-SB).

In the multi-label scenario, each input sequence has multiple target sequences with labels derived from different alphabet categories. $N$ categories $C=\{C_1, C_2, C_3, \cdots, C_N \}$ are then defined, where $C_i$ represents the alphabet of category $i$. For any input $x$ with target sequence $l$, each element in $l$ is then decomposed into $N$ components representing the $N$ categories where:

\begin{equation}
    \label{eq:MCTC1}
    l^t=(l^t_1, l^t_2, l^t_3, \cdots, l^t_N)
\end{equation}

Therefore, the input $x$ of length $T$ and target sequence $l$ of length $U \leq T$ can be represented in $N$ sequences of $l_i$ all of length $U$.

The network output tensor $y$ will be of size $T\times(\sum^N_{i=1}|C_i|+1)$ where $|C_i|$ is the number of elements in category $C_i$ plus the shared blank node. Let $C_i^{'}= C_i\cup{blank}$, then the probability of output element $j$ of category $C_i^{'}$ at time $t$ is denoted as $y_{i,j}^t$. In this case, the softmax function is applied over the components of $C_i^{'}$, hence $\sum_{j=1}^{|C_i^{'}|} y_{i,j}^{t}=1$.

The probability of each category is then computed separately as:

\begin{equation}
    \label{eq:MCTC2}
    p(l_i|x)=\sum_{\pi_{i}\in\beta_{i}^{-1}(l_i)}p(\pi_i|x)
\end{equation}

Where $\pi_i$ is a label sequence of length $T$ with $\pi_i^t\in C_i^{'}$ while $\beta_i$ is a many-to-one mapping defined for each category $i$ that maps $\pi_i$ to $l_i$, and the probability of the ouput path $\pi_i$ is computed as:

\begin{equation}
    \label{eq:MCTC3}
    p(\pi_i|x)=\prod^T_{t=1}y^t_{i,\pi^t_i}
\end{equation}

The final objective function is then computed as the multiplication of the label probabilities of all categories:

\begin{equation}
    \label{eq:MCTC4}
    p(l|x)=\prod_{i=1}^N p(l_i|x)
\end{equation}

\subsection{Speech Attribute Modeling}
$N = 35$ speech attributes were adopted representing the manners and places of articulation along with other phonological features as listed in Table \ref{tab:attrlist}. These attributes were selected so that each phoneme has a unique binary representation in terms of the 35 attributes. The 35 speech attributes were jointly learnt using the proposed SCTC-SB criterion. A category for each attribute ($att$) was defined with items $C_i={+att,-att}$. The category that represents the nasal attribute, for example, has possible outputs of {$+nasal,-nasal$}. Therefore, the number of network outputs is equal to 71, where 35 nodes represent the existence of each attribute ($+att$), 35 nodes represent the absence of each attribute ($-att$), and one node represents the blank output that is shared among all categories.

\begin{table}
\centering
\caption{\label{tab:attrlist} The 35 learned speech attributes including manners and places of articulation and other phonological features such as voiceness \cite{RN296}}
\begin{tabular}{m{4cm} m{4cm} m{4cm}}
 \hline
 \centering\textbf{Manners}  &  \centering\textbf{Places} & \centering\arraybackslash\textbf{Others} \\
 \hline
 Consonant, sonorant, fricative, nasal, stop, approximant, affricate, liquid, vowel, semivowel, continuant  & Alveolar, palatal, dental, glottal, labial, velar, mid, high, low, front, back, central, anterior, posterior, retroflex, bilabial, coronal, dorsal  & Long, short, monophthong, diphthong, round, voiced \\
 \hline
 \end{tabular}
\end{table}

\begin{figure}[h]
    \centering
    \includegraphics[width=0.9 \textwidth]{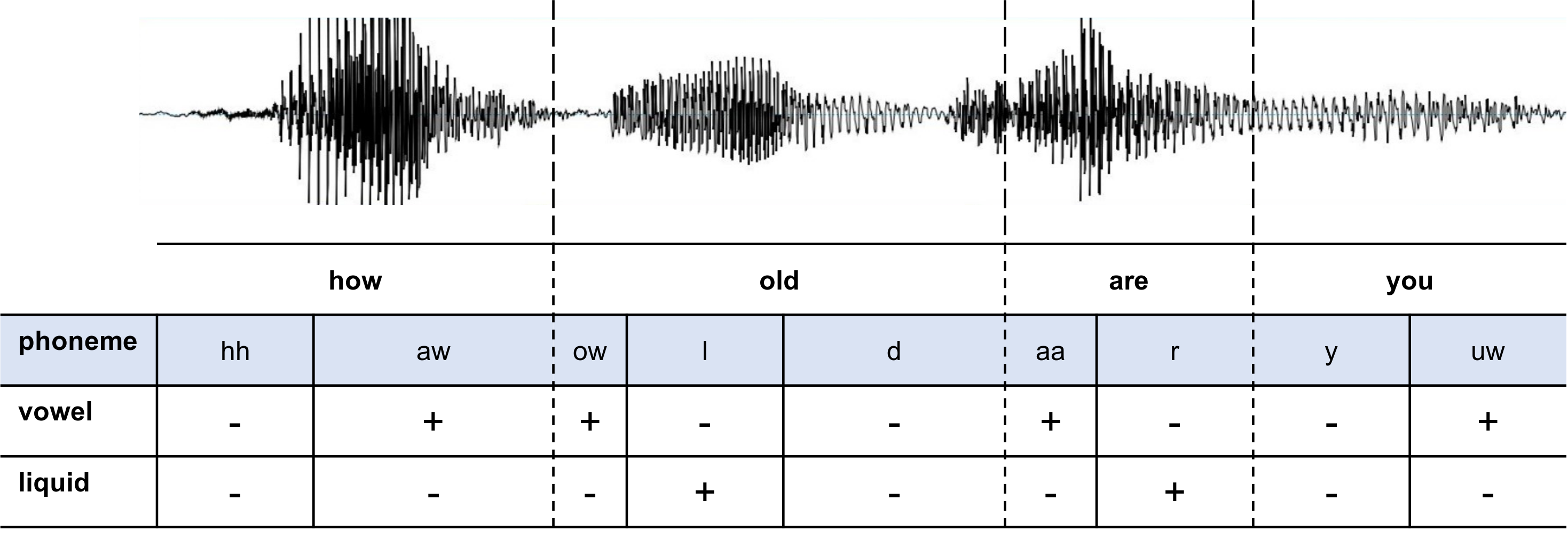}
    \caption{Example attribute mapping for the phrase ‘How old are you?’ showing the ground truth phoneme sequence and the corresponding vowel and liquid attributes labels. Here the vowel attribute is present for /aw/, /ow/,/aa/, and /uw/, while the liquid is present for /l/ and /r/ consonants.}
    \label{fig:method3}
\end{figure}

In the training phase, the phoneme sequence $l$ of each utterance was mapped to 35 target sequences  $l_i$ of $+att,-att$ symbols, one for each attribute as shown in Figure \ref{fig:method3} for an example phrase \emph{"How old are you?"}. The CTC loss of each category was then computed using (\ref{eq:MCTC3}) and the final loss function for the utterance was calculated according to (\ref{eq:MCTC4}) by multiplying the 35 CTC losses of all categories producing the phoneme sequence loss.

In the inference phase, for any speech input $x$ of length $T$, the most probable labelling of each attribute $i$ was obtained by applying an arg max function over the output nodes representing items in each category $C_i^{'}$ as follows

\begin{equation}
    \label{eq:att1}
    h_i(x)=\operatorname{argmax}_{j} y_{i,j}^{t},\ \ \ \ j=1,2,\cdots,|C_i^{'}|
\end{equation}

The output is a sequence of $+att,-att$ and blank tokens of length T. Finally, the repeated tokens were merged, and all blank tokens were removed to get the final output sequence of length $U \leq T$ using predefined category mapping $\beta_i$.

\section{Speech Corpora}
\label{sec:SpCor}
Three English speech corpora were employed in this paper. The native Librispeech (LS) \cite{RN39} corpus, which consists of audiobooks from the Librivox project \cite{RN294}, the native TIMIT dataset \cite{RN163}, and the non-native second language learning L2-ARCTIC corpus \cite{RN96}. The LS corpus is a frequently utilized open-access database for evaluating diverse speech-processing tasks. On the other hand, the TIMIT dataset is a well-designed, small dataset that includes phonetically balanced recordings from various dialects. Additionally, the non-native L2-ARCTIC corpus is a relatively recent speech corpus that has become the standard dataset for evaluating MDD systems.

The LS dataset has “clean” and “other” versions, referred to as LS-clean and LS-other, respectively. The “other” portion of the LS dataset, LS-other, contains challenging speech data that has lower recognition accuracy compared to the “clean” portion \cite{RN39}.

On the other hand, L2-ARCTIC was collected from 24 non-native English speakers equally distributed over 6 native languages, namely Arabic, Hindi, Korean, Mandarin, Spanish, and Vietnamese. The L2-ARCTIC contains two subsets, scripted and spontaneous, referred to as L2-Scripted and L2-Suitcase respectively. The scripted one consists of ~27 hours of speech, of which only ~3.5 hours were manually annotated at the phoneme level. In contrast, the spontaneous speech subset contains ~26 minutes of speech where each speaker recorded around one minute. Unlike the scripted speech, the whole spontaneous speech utterances were manually annotated at the phoneme level. The same speakers recorded both scripted and spontaneous subsets. The L2-ARCTIC was further split into training and testing subsets following the same split as in \cite{RN32}, where six speakers, one from each L1 language, formed the test set while the other 18 speakers were used for training.

\begin{table}[!h]
\centering
\caption{\label{tab:dataset1} Distribution of the utilised datasets.}
\begin{tabularx}{0.9\textwidth}{>{\centering\arraybackslash}X >{\centering\arraybackslash}X>{\centering\arraybackslash}X>{\centering\arraybackslash}X>{\centering\arraybackslash}X}
\cline{2-5}
 & Name & Type & Hours & Speakers \\ \hline\hline
\multirow{5}{*}{Training} & \cellcolor[HTML]{EFEFEF}TIMIT & \cellcolor[HTML]{EFEFEF}Native & \cellcolor[HTML]{EFEFEF}3.9 & \cellcolor[HTML]{EFEFEF}462 \\ \cline{2-5}
& LS-clean-100 & Native & 100 & 251 \\ \cline{2-5}
& \cellcolor[HTML]{EFEFEF}TIMIT+L2 & \cellcolor[HTML]{EFEFEF}Native+Non-Native & \cellcolor[HTML]{EFEFEF}6.5 & \cellcolor[HTML]{EFEFEF}480 \\ \hline\hline
\multirow{10}{*}{Testing} & LS-clean-test & Native & 4.1 & 40 \\ \cline{2-5}
& \cellcolor[HTML]{EFEFEF} LS-other-test & \cellcolor[HTML]{EFEFEF} Native & \cellcolor[HTML]{EFEFEF} 4.4 & \cellcolor[HTML]{EFEFEF} 33 \\ \cline{2-5}
& TIMIT-test & Native & 1.4 & 168 \\ \cline{2-5}
& \cellcolor[HTML]{EFEFEF} L2-Scripted & \cellcolor[HTML]{EFEFEF} Non-Native & \cellcolor[HTML]{EFEFEF} 0.11 & \cellcolor[HTML]{EFEFEF} 6 \\ \cline{2-5}
& L2-Suitcase & Non-Native & 0.87 & 6 \\ \hline
\end{tabularx}
\end{table}

The TIMIT corpus, 100 hours of the LS-clean, referred to as LS-clean-100, and a combination of TIMIT and L2-ARCTIC corpora, referred to as TIMIT+L2, were used separately for training, developing and testing of the speech attribute detection and the phoneme recognition acoustic models. While the L2-ARCTIC was for the evaluation of the MDD system.

Around 15\% of the L2-ARCTIC has manual annotations at the phoneme level describing each pronunciation error by indicating the pronounced phoneme, if pronounced, and if the error is an insertion (I), deletion (D), or substitution (S). Table \ref{tab:dataset2} shows the number of each type of error, along with the number of correctly pronounced phonemes (C), in the training and testing subsets of the L2-Scripted and L2-Suitcase datasets.

\begin{table}[!h]
\centering
\caption{\label{tab:dataset2} The number of Correct (C), Substituted (S), Inserted (I), and Deleted (D) phonemes in the L2ARCTIC speech corpus.}
\begin{tabularx}{0.9\textwidth}{>{\centering\arraybackslash}X >{\centering\arraybackslash}X>{\centering\arraybackslash}X>{\centering\arraybackslash}X>{\centering\arraybackslash}X>{\centering\arraybackslash}X}
\cline{2-6}
 & Subset & C & S & I & D \\ \hline\hline
\multicolumn{1}{c|}{} & \cellcolor[HTML]{EFEFEF}Training & \cellcolor[HTML]{EFEFEF}79864 & \cellcolor[HTML]{EFEFEF}10474 & \cellcolor[HTML]{EFEFEF}772 & \cellcolor[HTML]{EFEFEF}2437 \\ \cline{2-6} 
\multicolumn{1}{c|}{\multirow{-2}{*}{L2-Scripted}} & Testing & 28331 & 3198 & 214 & 939 \\ \hline
\multicolumn{1}{c|}{} & \cellcolor[HTML]{EFEFEF}Training & \cellcolor[HTML]{EFEFEF}5736 & \cellcolor[HTML]{EFEFEF}1032 & \cellcolor[HTML]{EFEFEF}55 & \cellcolor[HTML]{EFEFEF}290 \\ \cline{2-6} 
\multicolumn{1}{c|}{\multirow{-2}{*}{L2-Suitcase}} & Testing & 2333 & 438 & 28 & 137 \\ \hline
\end{tabularx}
\end{table}

The CMU dictionary \cite{RN268} was used to obtain the phonetic transcription of the LS orthographic transcription. To match the phoneme set of LS and L2-ARCTIC, the TIMIT phonemes were converted from 61 to 39 following the mapping proposed in \cite{RN281}.
However, /zh/ was not mapped to /sh/ as the former is \(+voiced\) while the latter is \(-voiced\) and merging them can confuse the \emph{voiced} attribute model. All silence labels were further removed leaving the silence frames to be handled by the blank label.

\section{Experimental Settings}
\label{sec:expset}

\subsection{Experiments}

First, we performed a number of experiments to assess the performance of our proposed speech attribute detection method followed by a comparison between the phonological (speech attribute)-level MDD and the phoneme-based MDD. We conducted the following experiments:
\begin{enumerate}
    \item To investigate the influence of the pre-trained model size and the domain of the speech corpora used in the pre-training process, we compared the performance of the three wav2vec2-based pre-trained models with respect to speech attribute recognition: wav2vec2-base, wav2vec2-large, and wav2vec2-large-robust.
    \item To explore the robustness of the the proposed speech attribute recognition approach, we compared its performance when applied to different domains (LS and TIMIT) as well as when used with out-of-domain data (native/non-native).
    \item To demonstrate the effectiveness of the proposed wav2vec2-based speech attribute detection model, we compared its performance to an existing baseline based on the Deep Speech 2.0 \cite{RN137} model.
    \item We finally assessed the effectiveness of our speech attribute recognition approach when used for Mispronunciation Detection and Diagnosis (MDD) by applying the method to L2 speech and comparing it to phoneme-level MDD.
\end{enumerate}
\subsection{Training Procedures}
\label{ssec:tran}

\begin{figure}[h!]
    \centering
    \includegraphics[width=0.9 \textwidth]{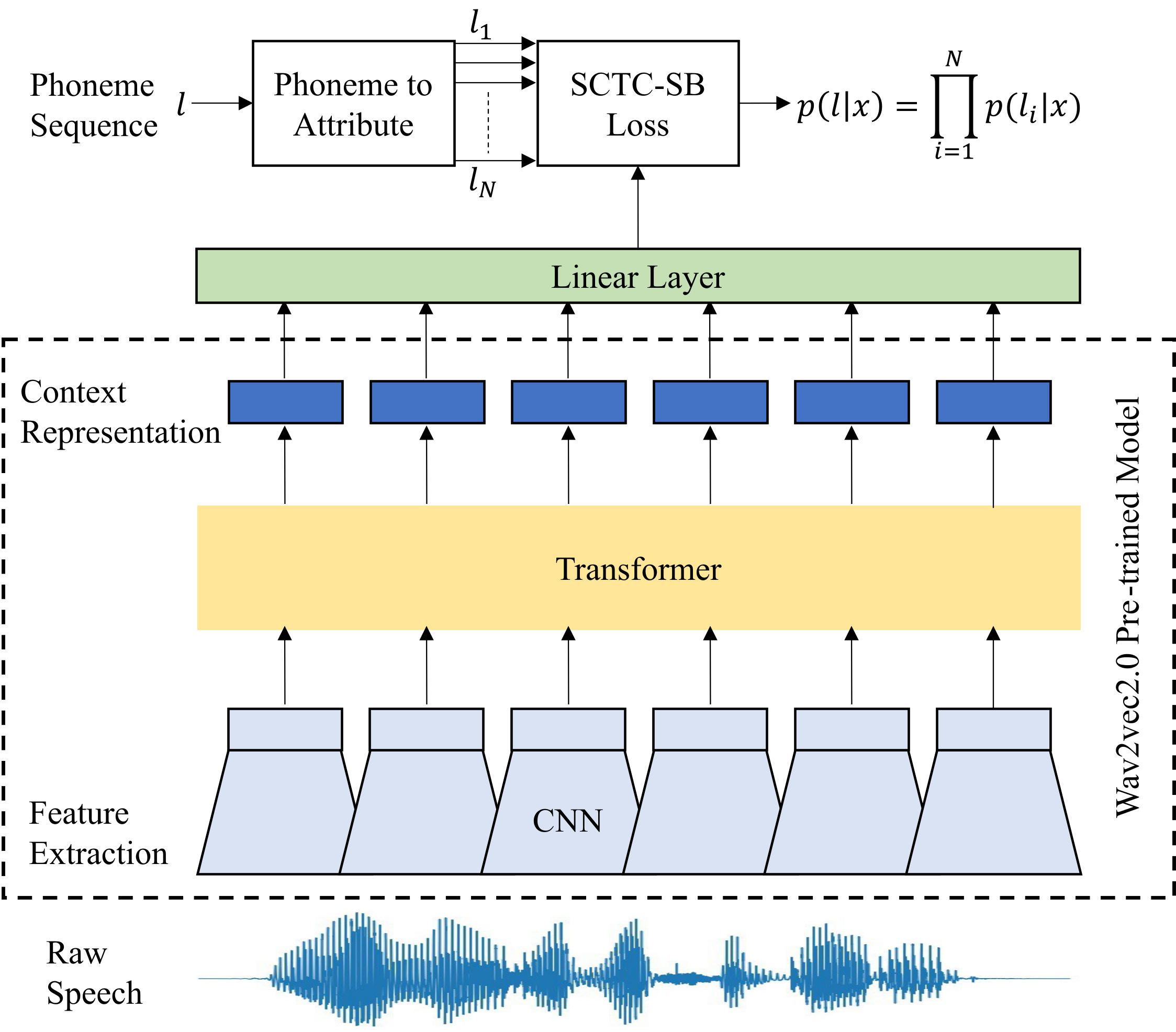}
    \caption{The training procedure for the proposed speech attribute recognition model. A linear layer was added on top of a wav2vec2-based pre-trained model. For each speech utterance, the linear layer converts the corresponding phoneme sequence to $N$ binary sequences of speech attributes. The CTC loss was then computed for each attribute sequence. Finally, the SCTC-SB loss was computed by multiplying all speech attributes’ CTC losses.}
    \label{fig:exp1}
\end{figure}

As aforementioned, the core model of the architecture is the self-supervised pre-trained wav2vec2 model. Figure \ref{fig:exp1} depicts the block diagram of the training procedure of our proposed model. The wav2vec2 model consists of a multi-layer CNN encoder that generates the latent representations followed by a transformer module with multiple blocks and multiple attention heads. A linear layer was added on top of the transformer module with the number of nodes representing the number of classes. The number of classes used to train the SCTC-SB-based speech attribute model was 71: 35 for the existence and 35 for the absence of each attribute plus one node for the \textit{blank} output. For the CTC-based phoneme recognition model, 40 output nodes were used representing the 39 phonemes in addition to a \textit{blank} node.

Except for the CNN encoder layer, the whole network was then fine-tuned to minimise either the SCTC-SB or CTC loss using backpropagation. As the feature extraction layer was already well-trained during pre-training, its parameters were fixed during the fine-tuning process. Furthermore, SpecAugment \cite{RN284} was applied to the output of the CNN encoder to add more variations to the training data.
The performance of three pre-trained models, namely wav2vec-base, wav2vec-large \cite{RN127}, and wav2vec-large-robust \cite{RN231} were compared. The first two models were pre-trained on 960 hours of read-out books dataset, Librispeech, while the latter one was pre-trained on multiple datasets from different domains, including read-out books and telephone conversations. The wav2vec-base model has 95M parameters while both wav2vec-large and wav2vec-large-robust have 317M parameters.

AdamW optimization \cite{RN285} was utilised for all experiments with a 0.005 weight decay and 0.0001 initial learning rate.  The batch size was fixed to 32, and the fine-tuning ran for 30 epochs. 10\% of the iteration steps were consumed in the warmup phase to reach the initial learning rate.

\subsection{Evaluation Metrics}
\label{ssec:eval}
In this work, we utilized several performance metrics to evaluate and compare the developed models based on their different tasks
\subsubsection{Attribute Recognition Performance}
For the attribute recognition model, as the output is a binary sequence of $+att/-att$ symbols, we used the traditional error rate derived from the Levenshtein distance metric \cite{RN295}. The Levenshtein distance metric works by measuring the difference between two sequences in terms of the number of Insertion (I), Deletion (D), and Substitution (S) edits. Therefore, the Attribute Error Rate (AER) was computed as follows:

\begin{equation}
    AER = \frac{S+D+I}{N}
\end{equation}

Where $I$, $D$, and $S$ were estimated by comparing the recognized sequence to the reference sequence and $N$ is the total number of reference symbols. In some experiments, we used Accuracy ($ACC$) instead of error rate which is simply computed as $1-Error Rate$.

Furthermore, for the speech attribute recognition task, we calculated the Precision ($PRE$), Recall ($REC$), and $F1$ score of each attribute as follows:

\begin{equation}
\begin{aligned}
    REC = TP/(TP+FN) \\ 
    PRE = TP/(TP+FP) \\ 
    F1 = 2\times \frac{PRE \times REC}{PRE+REC}
\end{aligned}
 \end{equation}

 Where $TP, FP$, and $FN$ are the True-Positive, False-Positive, and False-Negative, respectively. The positive and negative here refer to the generation of the $+att$ or $-att$ symbols.

 \subsubsection{Mispronunciation Detection and Diagnosis (MDD) Performance}
 For the evaluation of the MDD task, we used metrics as proposed in [63] at the phoneme-level. Firstly, we counted the occurrences of the following:

 \begin{enumerate}
     \item 	True-Acceptances ($TA$), which represent the number of times a recognized phoneme matches a correctly pronounced phoneme.
     \item True-Rejections ($TR$), which represent the number of times a recognized phoneme is different from a mispronounced phoneme.
     \item False-Acceptances ($FA$), which represent the number of times a recognized phoneme matches a mispronounced phoneme.
     \item False-Rejections ($FR$), which represent the number of times a recognized phoneme is different from a correctly pronounced phoneme.
 \end{enumerate}

The $TRs$ were further split to Correctly Diagnosed ($CD$) when the recognized phoneme matches the phoneme that was pronounced, and Diagnosis Error ($DE$) otherwise. We then used these counts to estimate the False Acceptance Rate ($FAR$), the False Rejection Rate ($FRR$), and the Diagnostic Error Rate ($DER$) as follows: 

\begin{equation}
\begin{aligned}
    FAR = FA/(FA+TR) \\ 
    FRR= FR/(FR+TA) \\ 
    DER= DE/(CD+DE)
\end{aligned}
\end{equation}

 Similar metrics were used for the attribute-level MDD. For instance, if the phoneme /s/ was mispronounced as /z/, this was considered a mispronunciation of the \textit{voiced} attribute only and correct pronunciation for all other attributes.

 \section{Results and Discussion}
 \label{sec:res}
 \subsection{Speech Attribute Recognition}
 \subsubsection{Comparison of Pre-Trained wav2vec2 Models}
 The goal of this experiment was to explore how the size of the model parameters and the nature of the training data impact the performance of speech attribute detection. Here, multi-label SCTC-SB speech attribute recognition models were trained by minimizing the SCTC-SB loss function in (\ref{eq:att1}) for all binary groups of the 35 speech attributes listed in Table \ref{tab:attrlist}. At inference time, a sequence of 35 $+att/-att$ tokens was produced for each speech file representing a sequence of the existence/absence of each speech attribute. The attribute recognition accuracy ($ACC$), precision ($PRE$), recall ($REC$) and F1 metrics were used for performance evaluation and model comparison.

 \begin{figure}[h]
    \centering
    \includegraphics[width= \textwidth]{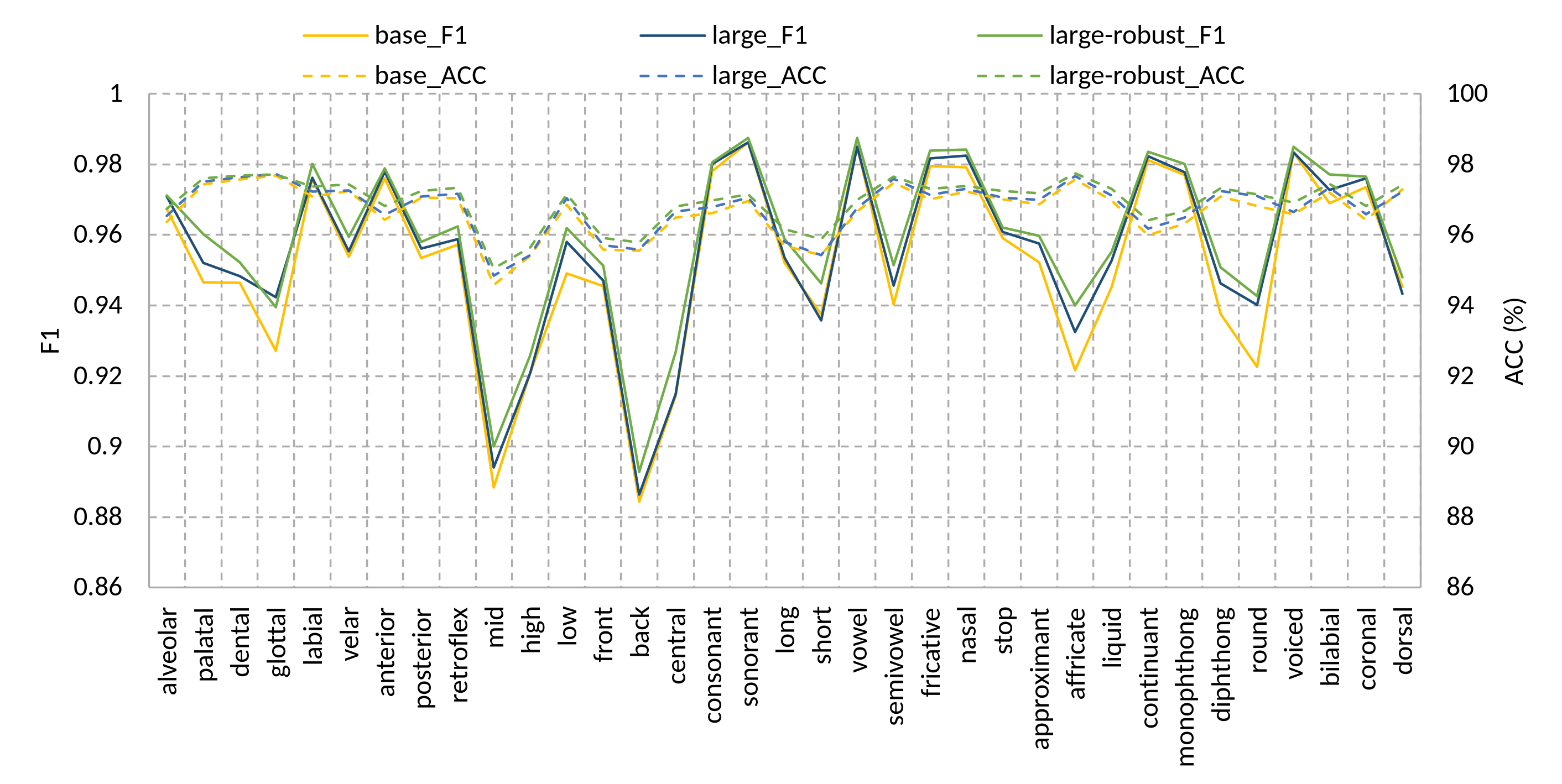}
    \caption{The performance of three wav2vec2-based pre-trained models for the speech attribute recognition task. The plots include the F1 (solid) and ACC (dashed) metrics (\%) of the 35 speech attributes for the TIMIT complete test set obtained from a model fine-tuned with the TIMIT training set. The base model has the lowest performance while the large-robust performed consistently higher for all attributes.}
    \label{fig:res1}
\end{figure}

Figure \ref{fig:res1} shows the $F1$ and $ACC$ metrics of the 35 speech attributes as recognised by the speech attribute recognition model when fine-tuned with the TIMIT training set and tested against the TIMIT test set. Here the performance of three wav2vec2-based pre-trained models, namely wav2vec2-base, wav2vec2-large, and wav2vec2-large-robust were compared.

Although both base and large models were pre-trained on the same dataset, the large model performed better than the base one. This demonstrates that increasing the model capacity improves the speech attributes recognition accuracy. On the other hand, the wav2vec2-large-robust model consistently slightly outperformed both base and large models for nearly all speech attributes. The average accuracies of the three models were $96.6\pm0.73$, $96.8\pm0.71$ and $97\pm0.65$ for the base, large and large-robust models, respectively. The large-robust model was pre-trained on data from different domains to be more robust against out-of-domain data \cite{RN231}. Similar behaviour was obtained when fine-tuning the three pre-trained models for the phoneme recognition task using the CTC criterion. The PERs of the base, large, and large-robust models were $8.1, 7.8$, and $7.3$ respectively.

In the subsequent experiments, the pre-trained wav2vec-large-robust model was used in the phoneme and speech attribute recognition downstream tasks.

\subsubsection{Performance of Speech Attribute Model Over Different Domains}

\begin{figure}[h]
    \centering
    \includegraphics[width= \textwidth]{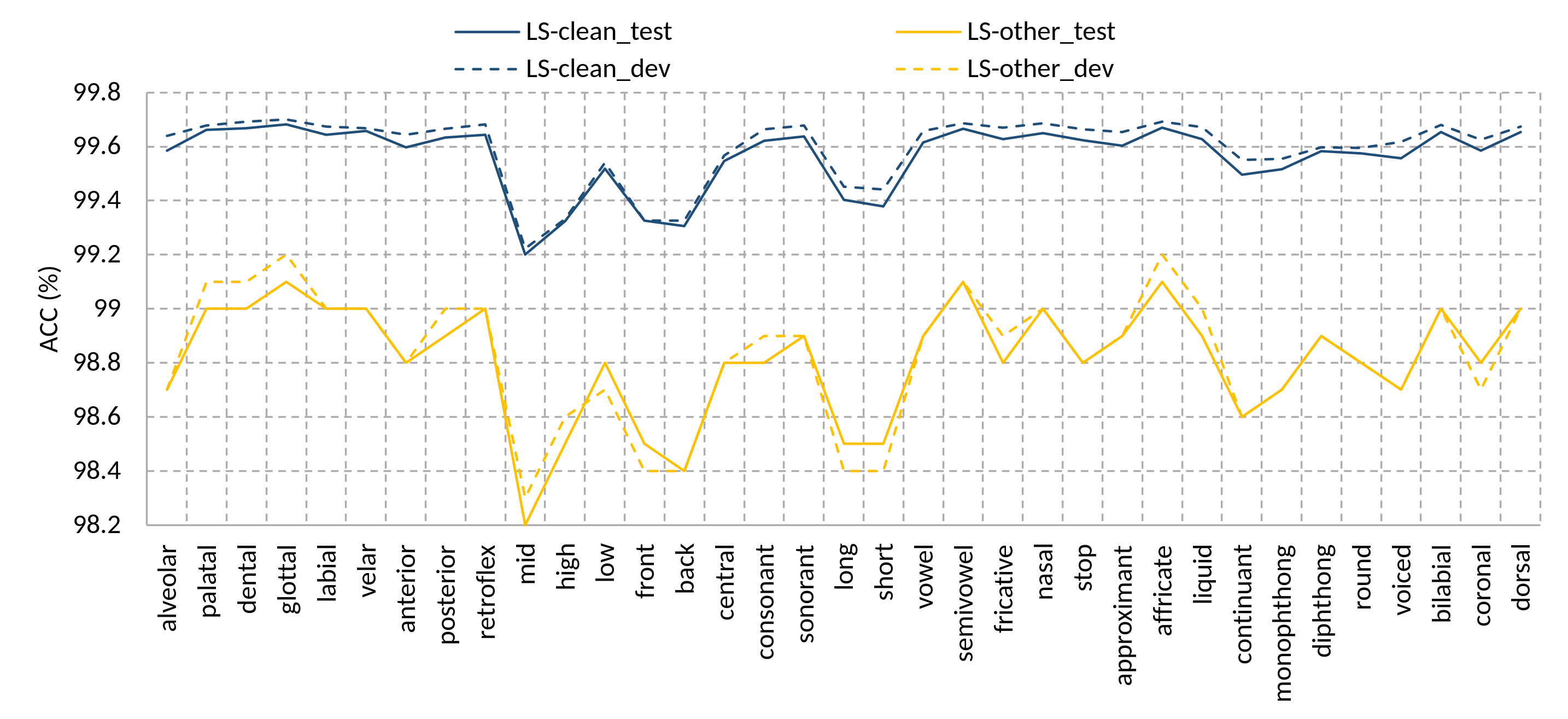}
    \caption{The 35 speech attribute recognition accuracies (\%) of the multi-label SCTC-SB model trained on the LS-clean-100 datasets and tested against the “clean” and “other” test and development sets. The “other” refers to a more challenging portion of the LS corpus. When applied to LS-other data, the model performance dropped consistently across all attributes with ratios ranging from ~31\% - ~45\%. Model performance declined the most for attributes associated with vowels, including \textit{high, front, mid}, and \textit{back}. On the other hand, attributes like \textit{fricative, stop, alveolar}, and \textit{dental} were less impacted.}
    \label{fig:res2}
\end{figure}

In this experiment, we investigated the robustness of the proposed model against challenging audio, like the "other" portion of the librispeech corpus (LS-other), and  in the case of domain shift from native to non-native speech. Figure \ref{fig:res2} shows the 35 speech attribute recognition accuracies of the proposed multi-label SCTC-SB model trained on the LS-clean-100 datasets and tested against the clean and other test and development sets, LS-clean-test, LS-clean-dev, LS-other-test, and LS-other-dev. The figure shows that the accuracies of all speech attributes of the clean test and development sets are above 99\%. However, applying the model to the more challenging LS-other speech datasets causes degradation across all speech attributes ranging from ~31\% up to ~45\%. The highest drop in the performance occurred in the vowel-related attributes such as \textit{high}, \textit{front}, \textit{mid} and \textit{back} while the \textit{fricative}, \textit{stop}, \textit{alveolar} and \textit{dental} were less affected attributes.

\begin{figure}[h]
    \centering
    \includegraphics[width= \textwidth]{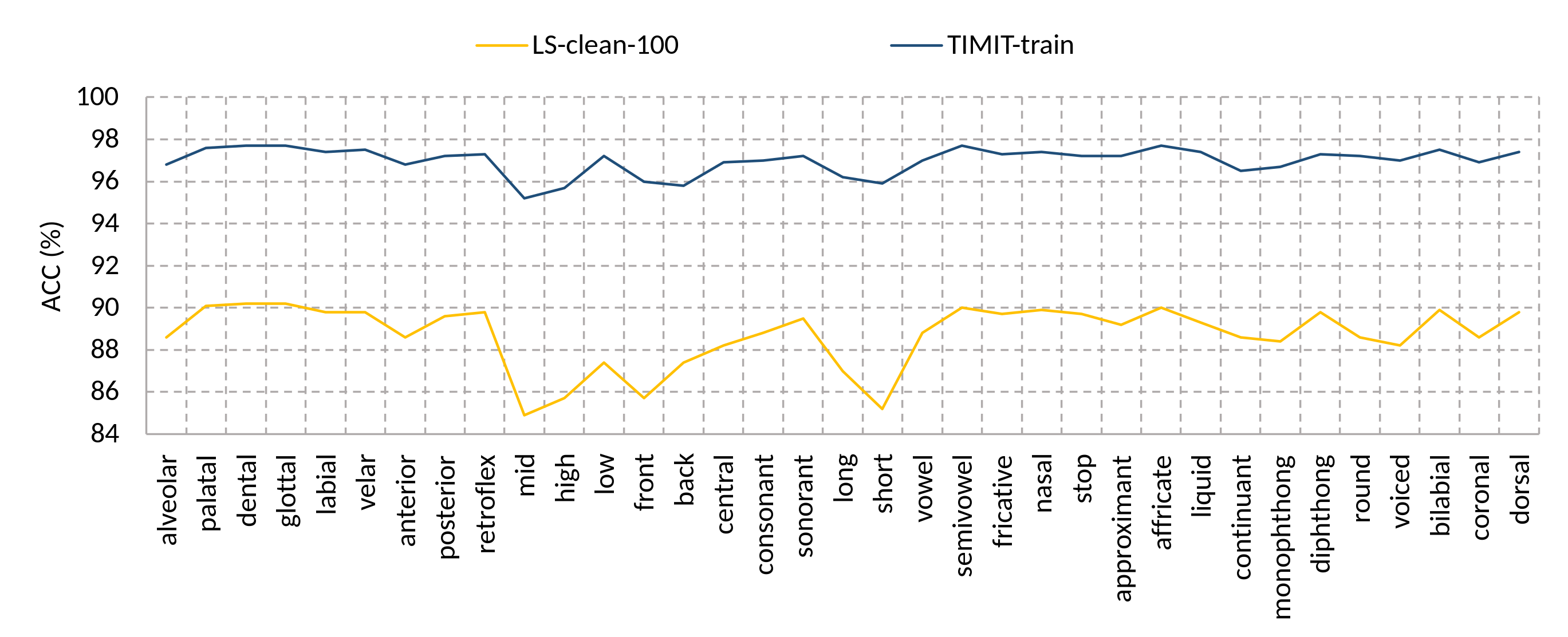}
    \caption{The effect of domain-mismatch between training and testing data. The plots show the ACC (\%) of the 35 speech attributes for multi-label SCTC-SB models trained on the LS-clean-100 and TIMIT training set and tested with the TIMIT test set. Testing the model with the out-of-domain data causes a relative degradation in the ACC ranging from ~22\% to ~33\%.}
    \label{fig:res3}
\end{figure}

To demonstrate the effect of the domain-mismatch between training and testing data, two multi-label SCTC-SB models were trained on LS-clean-100 and the TIMIT training set and tested against the TIMIT test set. As shown in Figure \ref{fig:res3}, the in-domain experiment, where the model was trained and tested on TIMIT speech corpus, achieved the lowest AER of around 2.3\% for dental, glottal, semivowel, and affricate attributes while the highest AER of 4.8\% was obtained by the mid attribute followed by high, back and short attributes. On the other hand, the model accuracy degradation ranged from ~22\% to ~33\% when trained on out-of-domain data, LS-clean-100.

\begin{figure}[h!]
    \centering
    \includegraphics[width= \textwidth]{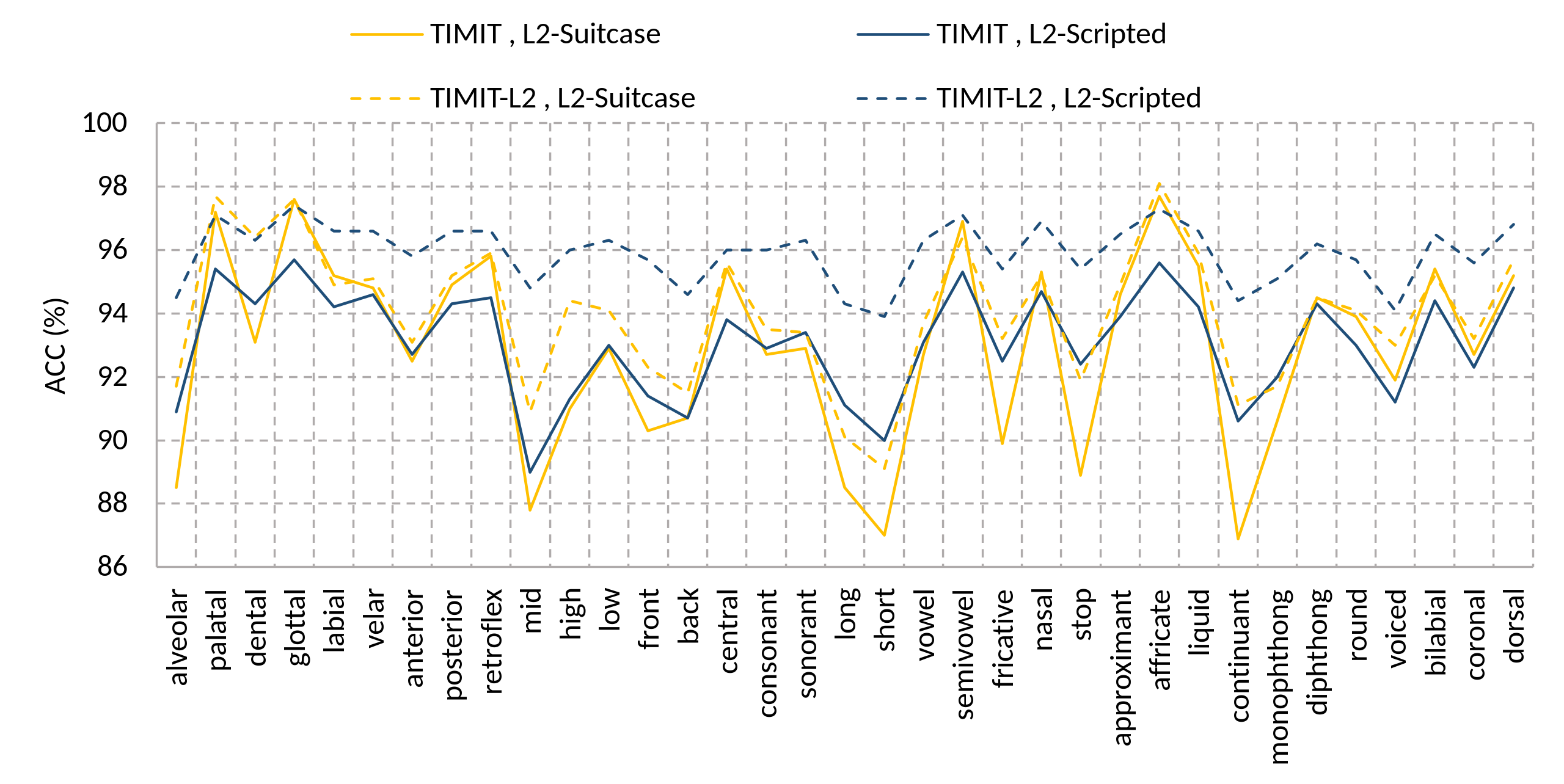}
    \caption{The accuracy of speech attributes recognition when used with non-native test sets. Here multi-label SCTC-SB models are fine-tuned with native (TIMIT) and native+non-native (TIMIT-L2) datasets and tested with non-native L2-Scripted and L2-Suitcase datasets. The L2-Scripted data achieves significant improvements when adding L2 data to the training set over all attributes.}
    \label{fig:res4}
\end{figure}

Figure \ref{fig:res4} shows how the speech attributes recognition accuracy of models fine-tuned on native (TIMIT) and native+non-native (TIMIT-L2) datasets were impacted when tested on non-native sets, L2-Scripted and L2-Suitcase. For the L2- Scripted test set, it is evident that adding non-native data to the training set significantly improves the recognition accuracy of all speech attributes. However, for L2-Suitcase, not all attributes show the same improvement when using non-native data in the model’s fine-tuning. The glottal, retroflex, nasal and diphthong attributes achieved almost the same accuracy with and without non-native data. Moreover, accuracies of labial and semivowel degraded slightly with non-native data.

\begin{figure}[h!]
     \centering
     \begin{subfigure}[b]{0.32\textwidth}
         \centering
         \includegraphics[width=\textwidth]{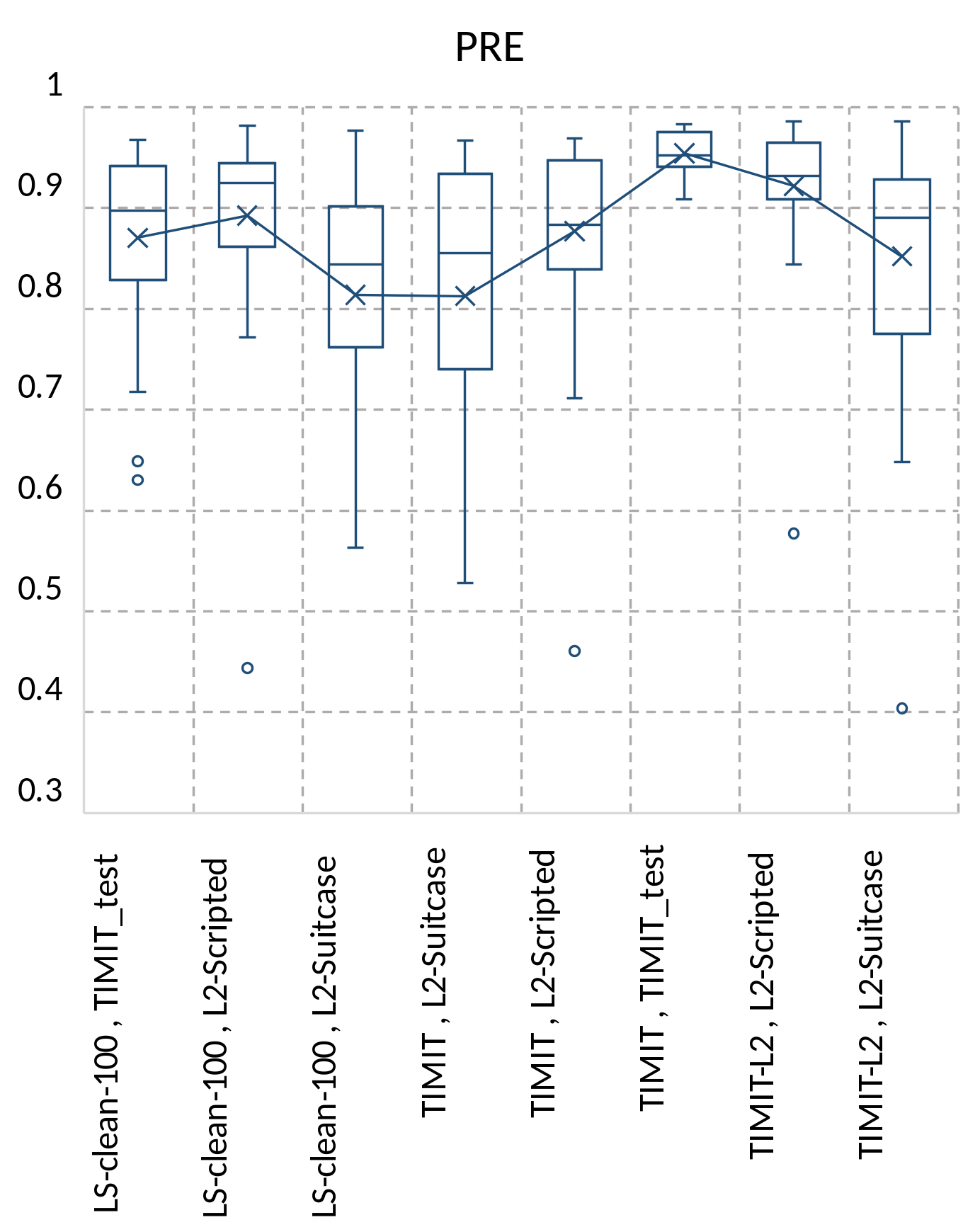}
     \end{subfigure}
     \hfill
     \begin{subfigure}[b]{0.32\textwidth}
         \centering
         \includegraphics[width=\textwidth]{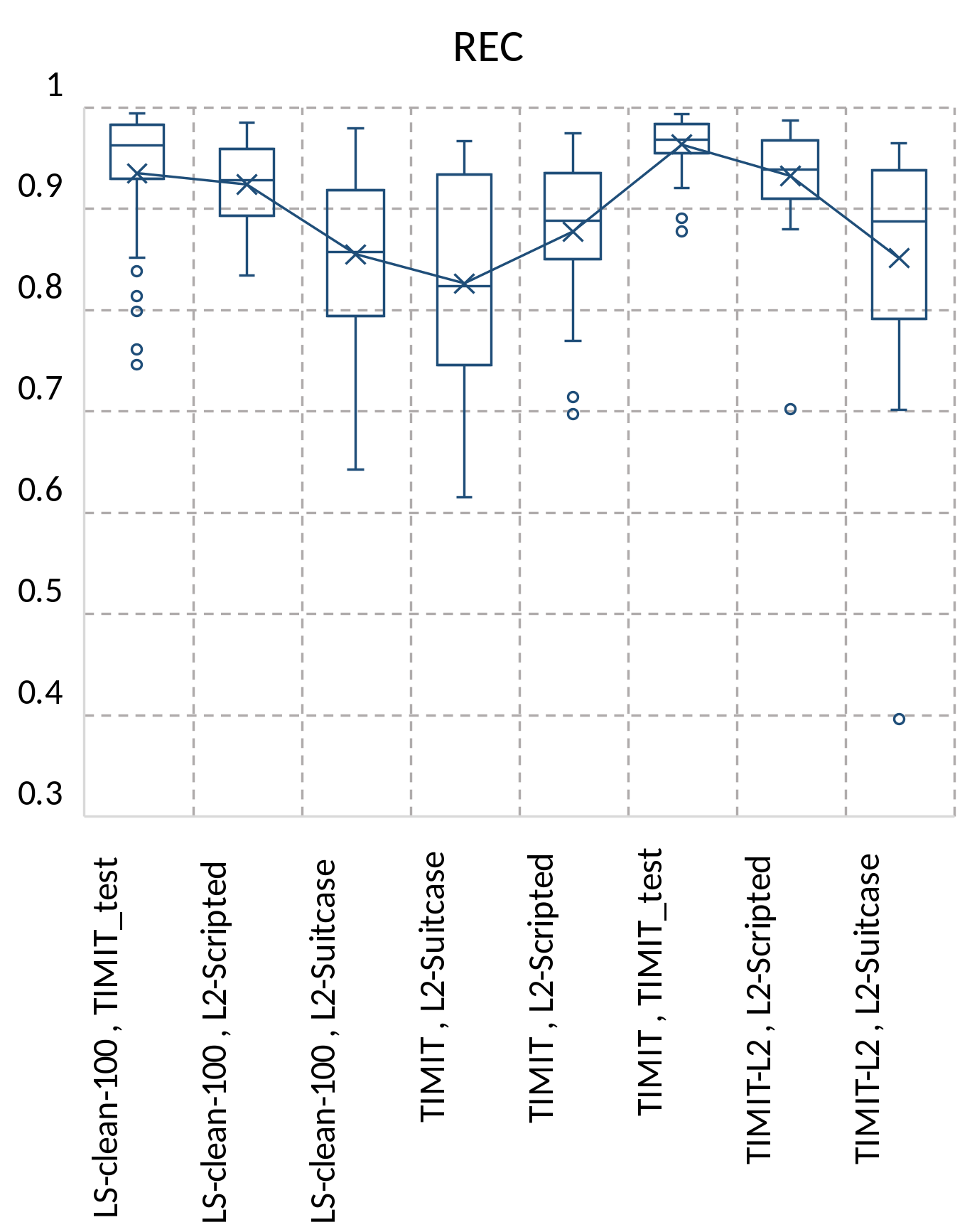}
     \end{subfigure}
     \hfill
     \begin{subfigure}[b]{0.32\textwidth}
         \centering
         \includegraphics[width=\textwidth]{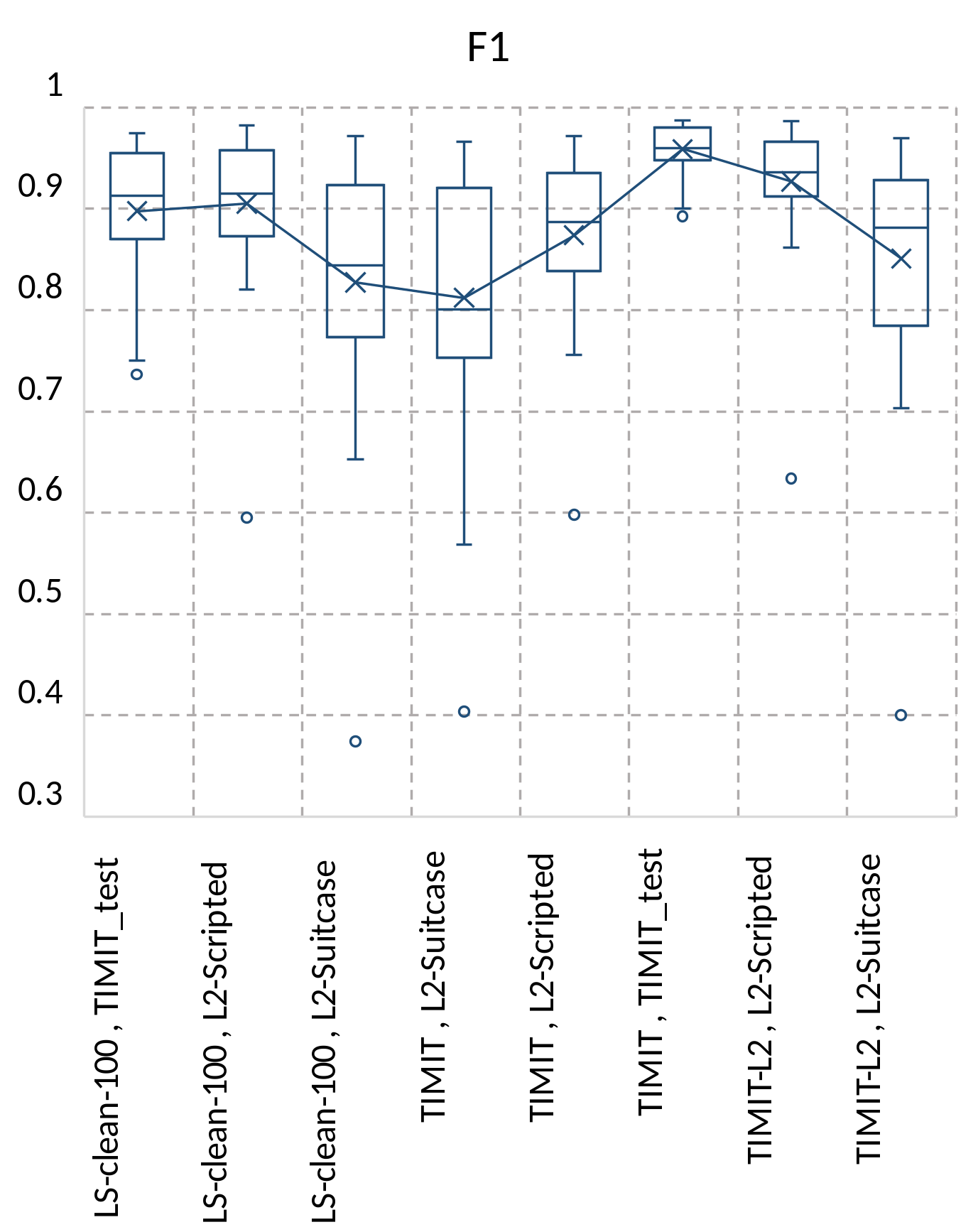}
     \end{subfigure}
        \caption{A comparison of the speech attribute recognition performance metrics across different pairs of training and testing datasets. Both TIMIT and LS-clean-100 are native English data, while L2-Scripted and L2-Suitcase are non-native English scripted and spontaneous speech, respectively. The boxplots show the precision (PRE), recall (REC), and F1 quartiles of the different (training, testing) data pairs. The spontaneous L2-Suitcase test set achieves lower PRE, REC, and F1 compared with the L2-Scripted test set over all training sets.}
        \label{fig:res5}
\end{figure}

Figure \ref{fig:res5} summarises the precision, recall, and F1 quartiles of different pairs of training and testing sets. It is noticeable that the performances of both TIMIT and L2-Scripted with a multi-label SCTC-SB model trained on LS-clean-100 were comparable despite both TIMIT and LS-clean-100 being native English data. Moreover, although both L2-suitcase and L2-Scripted were recorded by the same speakers, the L2-Suitcase seems more challenging than the L2-Scripted. This is explained by the L2-Suitcase recordings' longer duration and spontaneous nature compared to the short and scripted L2-Scripted dataset.

\subsubsection{Comparison With Existing Work} \label{sssec:Comp}

Most of the existing work on speech attribute detection reports frame-level performance. However, the proposed model is based on CTC, which outputs a sequence of symbols. As the CTC blank node allows the model to output no label in some frames when uncertain, obtaining frame-level alignment from the CTC model is not accurate. Consequently, it is hard to make a fair comparison with a frame-level attribute detection system.
Therefore, the proposed SCTC-SB model was compared with two CTC-based systems. Both systems were based on the Deep Speech 2 model \cite{RN137} and trained using CTC. The first system was proposed in \cite{RN162} for nasality detection where the output is a sequence of +nasal,-nasal tokens. The model achieved an error rate of 4.4\%. The second system, introduced in \cite{RN169}, discriminates between five manners of articulations, namely vowel, semi-vowel, nasal, stop, and fricative. The system achieved an error rate of 2.7\%. Both systems were trained and tested on the LS-clean-100 and LS-clean datasets, respectively. As shown in Figure \ref{fig:res3}, when the proposed SCTC-SB system was trained and tested on the same dataset, the error rate for all attributes was less than 1\%, significantly outperforming both baseline systems.

\subsection{Mispronunciation Detection and Diagnosis (MDD)}
The next experiments aim to demonstrate the effectiveness of the proposed speech attribute-based MDD in terms of detection and diagnosis accuracies. For comparison, we implemented a SOTA phoneme-level MDD and applied it to the same native and non-native speech corpora.
\subsubsection{Phoneme-level MDD}
In this experiment, MDD at the phoneme level was performed using three phonetic acoustic models trained on the LS-clean-100, TIMIT, and TIMIT-L2 datasets. Table \ref{tab:res1} summarises the evaluation results of the different models tested with L2-Scripted and L2-Suitcase test sets. The table shows that the model fine-tuned with the LS-clean-100 dataset has the highest $FAR$s of ~63\% and ~57\% for L2-Scripted and L2-Suitcase respectively, and an $FRR$ of ~8\% for both datasets. In contrast, the model fine-tuned with the TIMIT dataset achieved a better balance between $FRR$ and $FAR$. Adding L2 data to the TIMIT training dataset significantly reduced the $FRR$ from ~18\% and ~21\%, when using only the TIMIT training set, to ~7\% and ~12\% for the L2-Scripted and L2-Suitcase test sets, respectively. Furthermore, the model fine-tuned with the TIMIT-L2 dataset achieved the lowest $DER$s of ~15\% and ~17\% for L2-Scripted and L2-Suitcase test sets, respectively. These results suggest that by increasing the amount of native data in the training of the phonetic acoustic model, the model’s tendency to accept pronunciation errors ($FAR$) is increased significantly. By comparing the TIMIT and TIMIT-L2 models, it is noticeable that although both models can detect the existence of error with a comparable $FAR$, adding in-domain L2 data improved the ability of the model to Correctly Diagnose ($CD$) the error and significantly reduce the $DER$.

\begin{table}[!h]
\centering
\caption{\label{tab:res1} Evaluation results for the non-native test data using phonetic acoustic models trained on native and non-native datasets for a phoneme-level MDD task. Using a combination of native and non-native training data achieved the lowest DER over the two test sets.}
\begin{tabularx}{\textwidth}{>{\centering\arraybackslash\hsize=2.3\hsize}X||>{\centering\arraybackslash\hsize=1.3\hsize}X>{\centering\arraybackslash\hsize=0.7\hsize}X>{\centering\arraybackslash\hsize=0.7\hsize}X>{\centering\arraybackslash\hsize=0.8\hsize}X>{\centering\arraybackslash\hsize=0.8\hsize}X>{\centering\arraybackslash\hsize=0.8\hsize}X>{\centering\arraybackslash\hsize=0.9\hsize}X>{\centering\arraybackslash\hsize=0.9\hsize}X>{\centering\arraybackslash\hsize=0.8\hsize}X}
\hline
Train Data & Test Data & FA & FR & TA & \multicolumn{2}{c}{TR} & FRR (\%) & FAR (\%) & DER (\%) \\
\hhline{~||~~~~==~~~}
&&&&&CD&DE&&& \\
\hline\hline
\multirow{2}{*}{LS-clean-100} & L2-Scripted & 2686 & 2261 & 23920 & 1077 & 497 & 8.64 & 63.05 & 31.58 \\
& \cellcolor[HTML]{EFEFEF} L2-Suitcase & \cellcolor[HTML]{EFEFEF} 345 & \cellcolor[HTML]{EFEFEF} 256 & \cellcolor[HTML]{EFEFEF} 1884 & \cellcolor[HTML]{EFEFEF} 191 & \cellcolor[HTML]{EFEFEF} 66 & \cellcolor[HTML]{EFEFEF} 11.96 & \cellcolor[HTML]{EFEFEF} 57.31 & \cellcolor[HTML]{EFEFEF} 25.68 \\
\hline\hline
\multirow{2}{*}{TIMIT} & L2-Scripted & 1649 & 4808 & 21300 & 1896 & 715 & 18.42 & 38.71 & 27.38 \\
& \cellcolor[HTML]{EFEFEF} L2-Suitcase & \cellcolor[HTML]{EFEFEF} 209 & \cellcolor[HTML]{EFEFEF} 455 & \cellcolor[HTML]{EFEFEF} 1649 & \cellcolor[HTML]{EFEFEF} 264 & \cellcolor[HTML]{EFEFEF} 129 & \cellcolor[HTML]{EFEFEF} 21.63 & \cellcolor[HTML]{EFEFEF} 34.72 & \cellcolor[HTML]{EFEFEF} 32.82 \\
\hline\hline
\multirow{2}{*}{TIMIT-L2} & L2-Scripted & 1683 & 1899 & 24079 & 2170 & 407 & \textbf{7.31} & \textbf{39.51} & \textbf{15.79} \\
& \cellcolor[HTML]{EFEFEF} L2-Suitcase & \cellcolor[HTML]{EFEFEF} 155 & \cellcolor[HTML]{EFEFEF} 268 & \cellcolor[HTML]{EFEFEF} 1829 & \cellcolor[HTML]{EFEFEF} 370 & \cellcolor[HTML]{EFEFEF} 77 & \cellcolor[HTML]{EFEFEF} \textbf{12.78} & \cellcolor[HTML]{EFEFEF} \textbf{25.75} & \cellcolor[HTML]{EFEFEF} \textbf{17.23} \\
\hline
\end{tabularx}
\end{table}

\subsubsection{Speech Attribute-level MDD}
In this experiment, the three datasets, LS-clean-100, TIMIT, and TIMIT-L2 were used to train three SCTC-SB speech attributes models to detect the existence or absence of 35 speech attributes listed in Table \ref{tab:attrlist}. The output of each model was a sequence of $+att/-att$ for each input speech signal. The evaluation was performed for each attribute separately by aligning each output sequence with the corresponding reference sequence obtained by mapping the target phoneme sequence to the target attribute sequence. The $FAR$, $FRR$, and $DER$ were computed as explained in section \ref{ssec:eval}.

\begin{figure}[h]
    \centering
    \includegraphics[width= \textwidth]{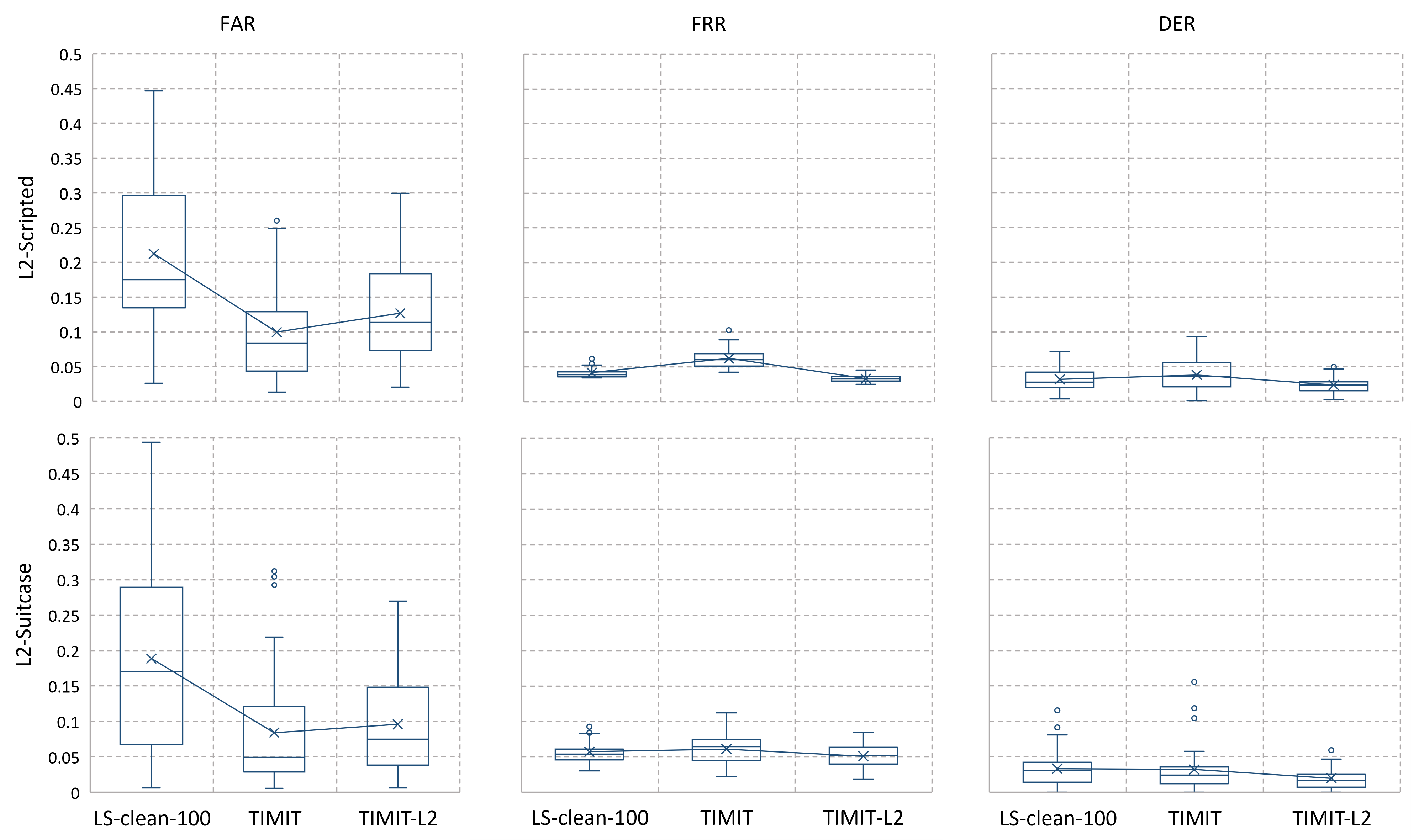}
    \caption{Evaluation of three speech attributes models in performing speech attribute level MDD of two test sets, L2-Scripted and L2-Suitcase. Two models were trained on native speech corpora, namely LS-clean-100, and TIMIT, while the third model was trained on a combination of native and non-native data TIMIT-L2.}
    \label{fig:res6}
\end{figure}

Figure \ref{fig:res6} depicts boxplots summarizing the $FAR$, $FRR$, and $DER$ of the 35 speech attributes MDD obtained using models trained on the three mentioned training sets and tested using the L2-Scripted and L2-Suitcase test sets. The results show that the $FAR$ of most speech attributes was less than 30\% when using the purely native model trained on LS-clean-100 which is significantly lower than the phoneme-level counterpart $FAR$s of ~57\% and ~63\% for L2-Suitcase and L2-Scripted respectively (see Table \ref{tab:res1}). The $DER$s of the three models tested with L2-Scripted for all speech attributes lies below 10\% which is also significantly lower than the phoneme-level $DER$s of ~31\%, ~27\%, and ~15\% of the LS-clean-100, TIMIT, and TIMIT-L2 respectively.

\begin{figure}[h!]
     \centering
     \begin{subfigure}[b]{\textwidth}
         \centering
         \includegraphics[width=\textwidth]{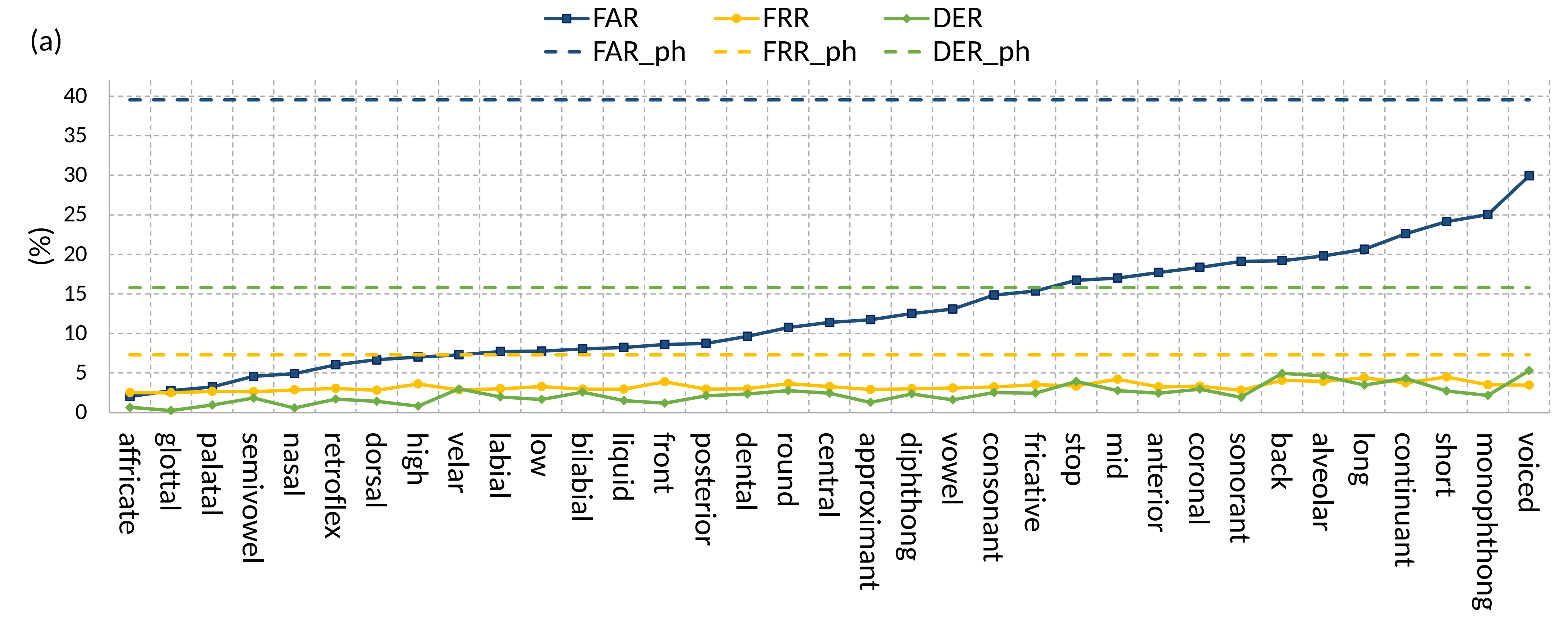}
     \end{subfigure}
     
     \begin{subfigure}[b]{\textwidth}
         \centering
         \includegraphics[width=\textwidth]{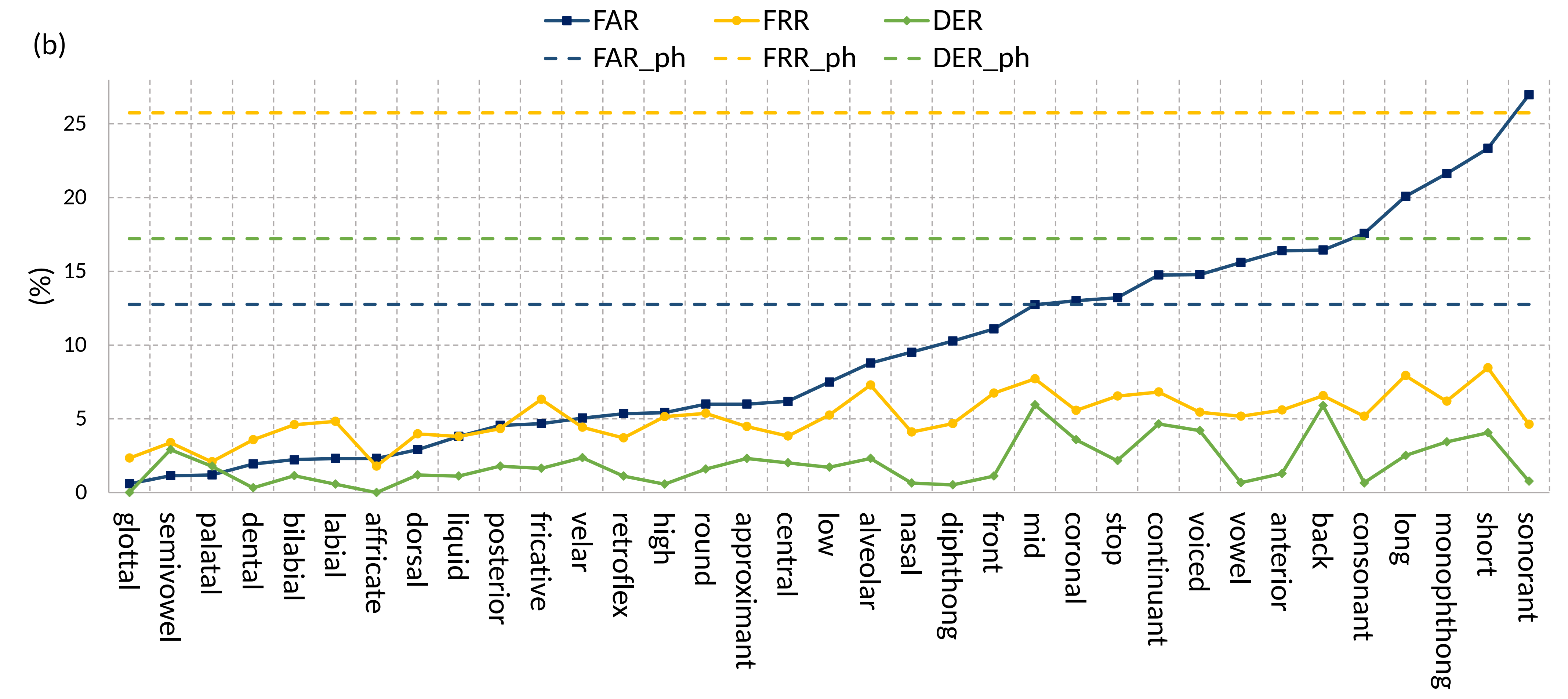}
     \end{subfigure}
        \caption{The speech attribute-level MDD performance of (a) L2-Scripted and (b) L2-Suitcase. The attributes are sorted from lowest to highest FAR. The straight lines FAR\_ph, FRR\_ph and DER\_ph are the phoneme-level MDD metrics. For L2-scripted the FAR for all attributes detected by the attribute level MDD was lower than the phoneme level MDD. The DER of the scripted test set (L2-Scripted) was consistently lower than that of the spontaneous test set (L2-Suitcase).}
        \label{fig:res7}
\end{figure}

Detailed results for each attribute obtained from the TIMIT-L2 model are shown in Figure \ref{fig:res7} for both L2-Scripted and L2-Suitcase test sets. The speech attributes are sorted from lowest to highest $FAR$ values. It is obvious that there are high variations in the $FAR$ among speech attributes while the $DER$ and $FRR$ are more consistent. The results demonstrate also that in both L2 test sets, all the speech attributes have achieved $FAR$, $FRR$, and $DER$ lower than the phoneme level equivalents (shown as straight lines).

\begin{figure}[h!]
    \centering
    \includegraphics[width= \textwidth]{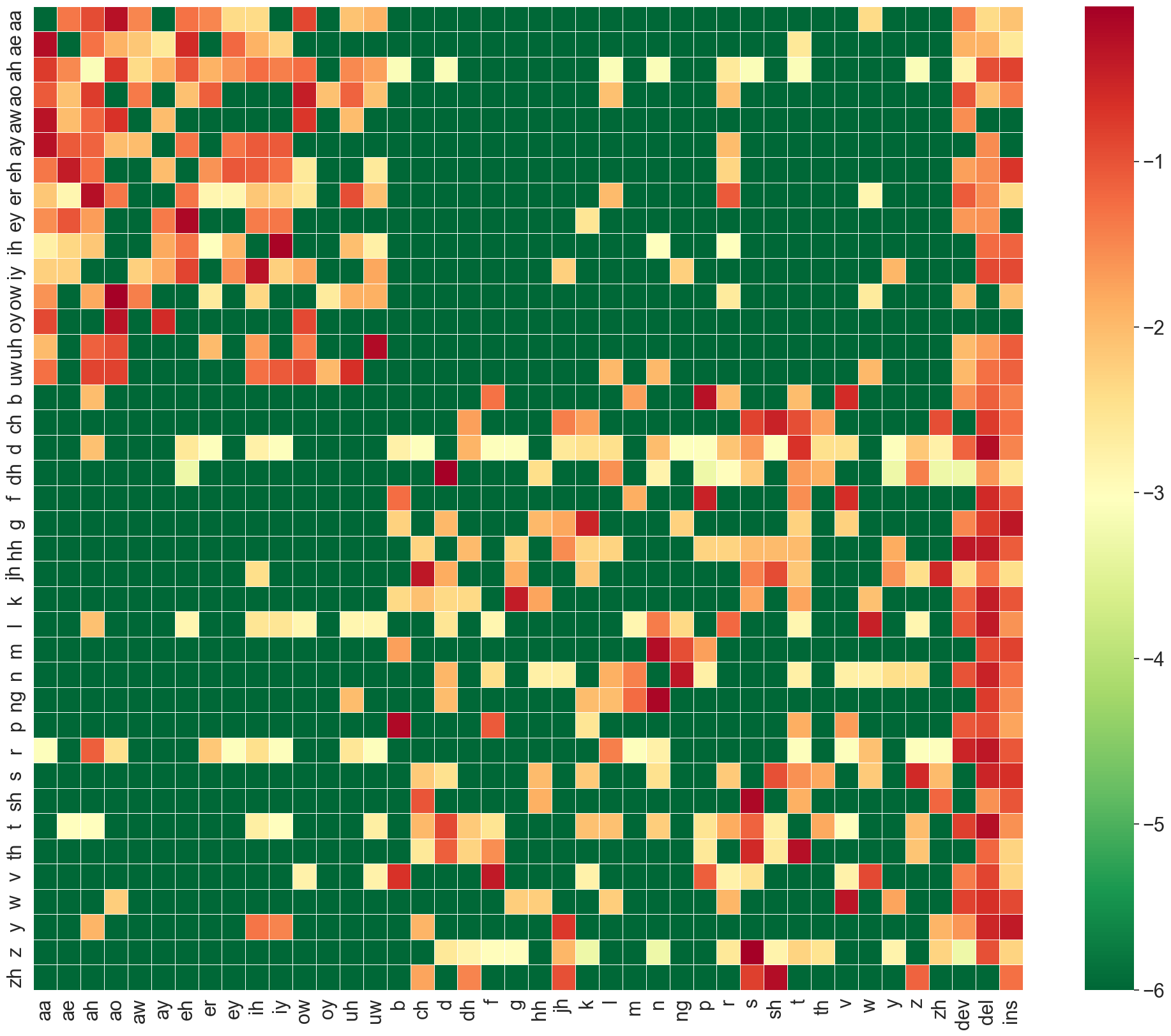}
    \caption{Pronunciation errors made by L2 speakers of L2-ARCTIC speech corpora. The vertical axis labels show the canonical phoneme, while the horizontal axis labels show the replaced phoneme, in case of substitution error, the deviation (dev), insertion (ins), or deletion (del). The colour represents the percentage of each error relative to the total number of errors of each phoneme on a log scale with red indicating the highest confusion rate and green indicating no confusion. There is a notable occurrence of both deletion and insertion errors across all sounds, particularly consonants. Additionally, within the category of consonants, there is a significant level of confusion between voiced and unvoiced sounds, such as /s/ and /z/, /p/ and /b/, as well as /d/ and /t/.}
    \label{fig:res8}
\end{figure}

To better understand the benefits and limitations of the speech attribute model in MDD, the phoneme-based and speech attribute-based models were used to detect 30 common substitution errors obtained from the manual annotation of the L2-Scripted dataset. A complete pronunciation error matrix for L2 speakers is depicted in Figure \ref{fig:res8}.

\begin{figure}[h!]
     \centering
     \begin{subfigure}[b]{0.48\textwidth}
         \centering
         \setlength{\fboxrule}{2pt}
         \includegraphics[width=\textwidth]{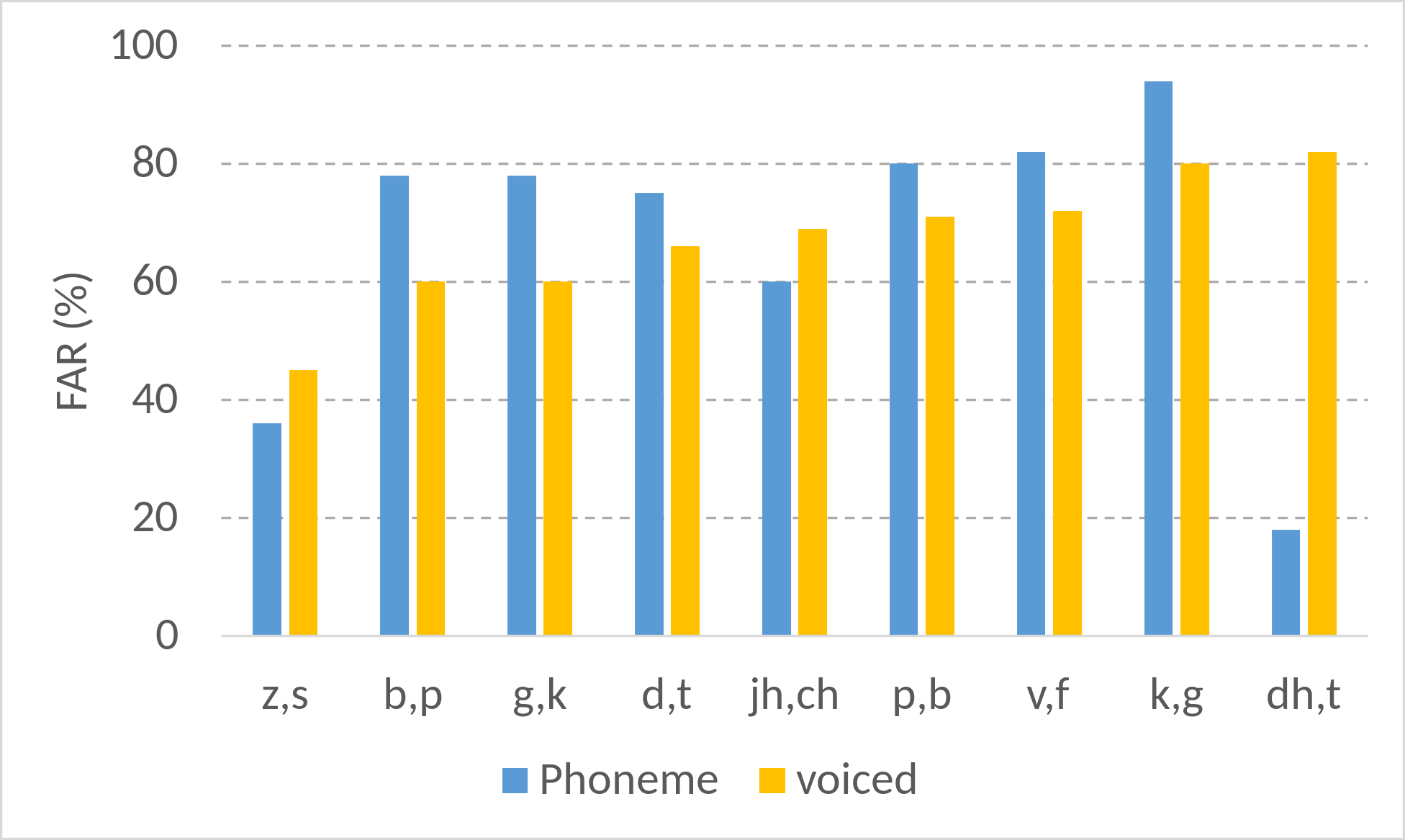}
     \end{subfigure}
     \hfill
     \begin{subfigure}[b]{0.48\textwidth}
         \centering
         \includegraphics[width=\textwidth]{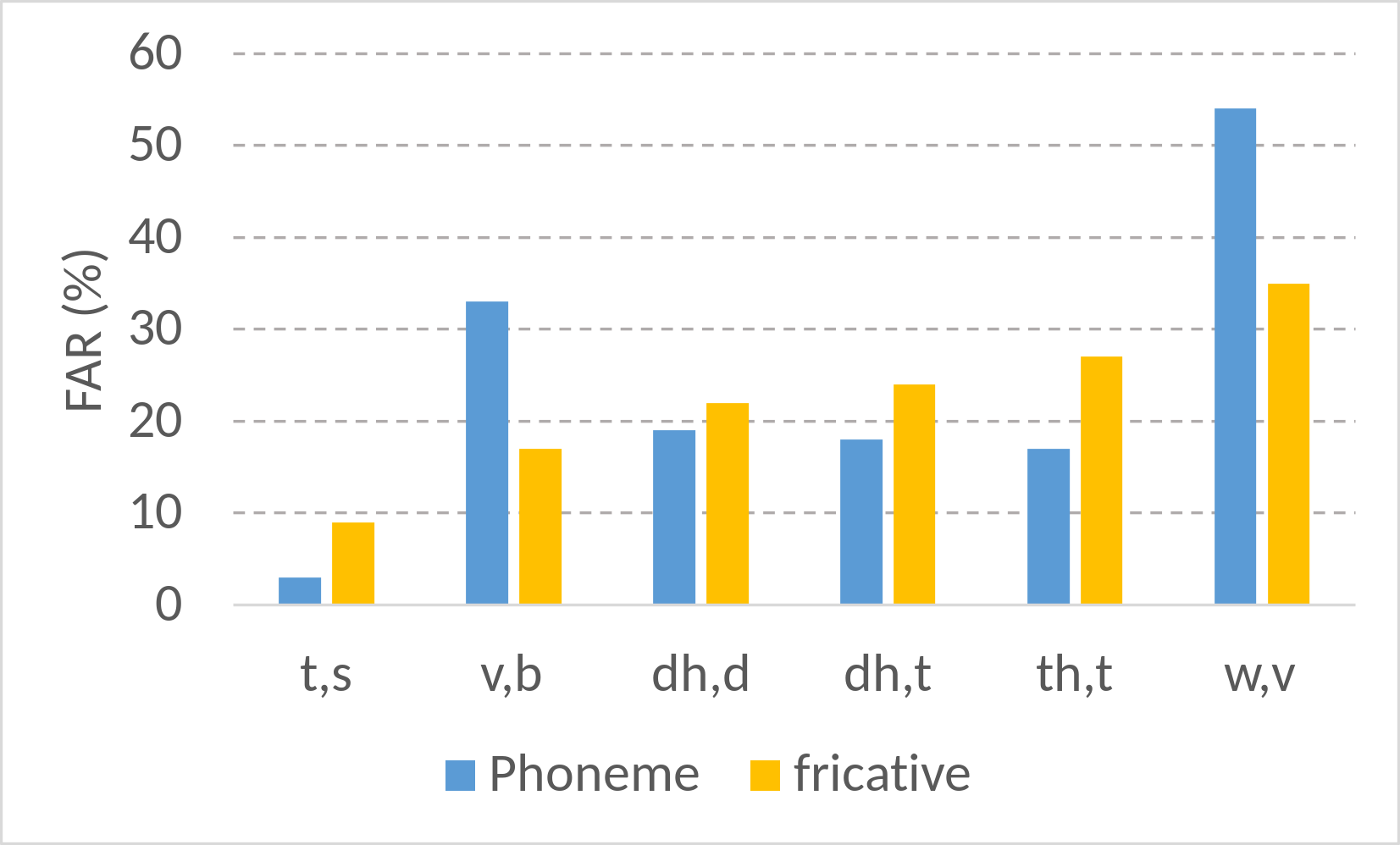}
     \end{subfigure}
     \begin{subfigure}[b]{0.48\textwidth}
         \centering
         \includegraphics[width=\textwidth]{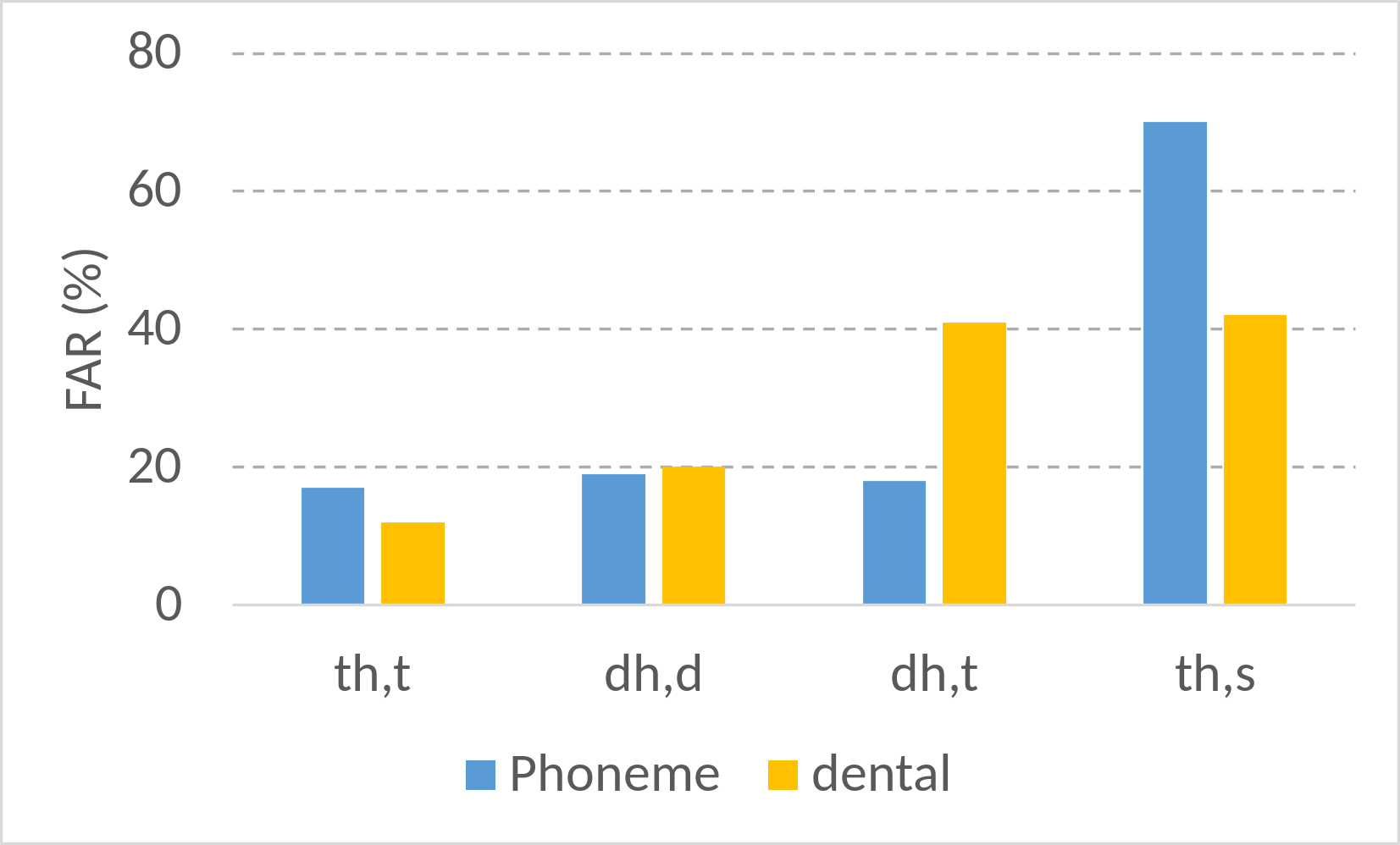}
     \end{subfigure}
     \hfill
    \begin{subfigure}[b]{0.48\textwidth}
         \centering
         \includegraphics[width=\textwidth]{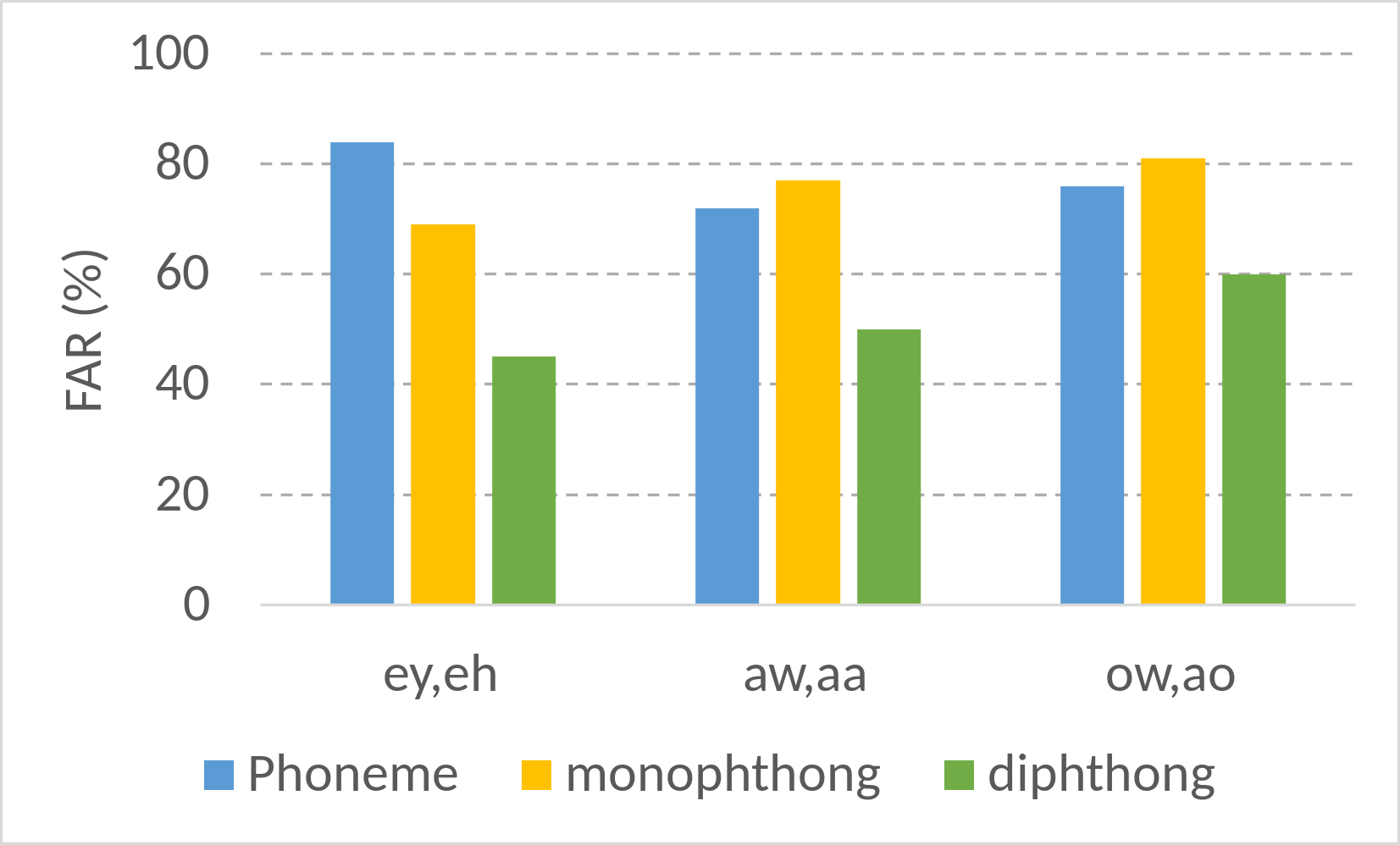}
     \end{subfigure}
     \begin{subfigure}[b]{0.48\textwidth}
         \centering
         \includegraphics[width=\textwidth]{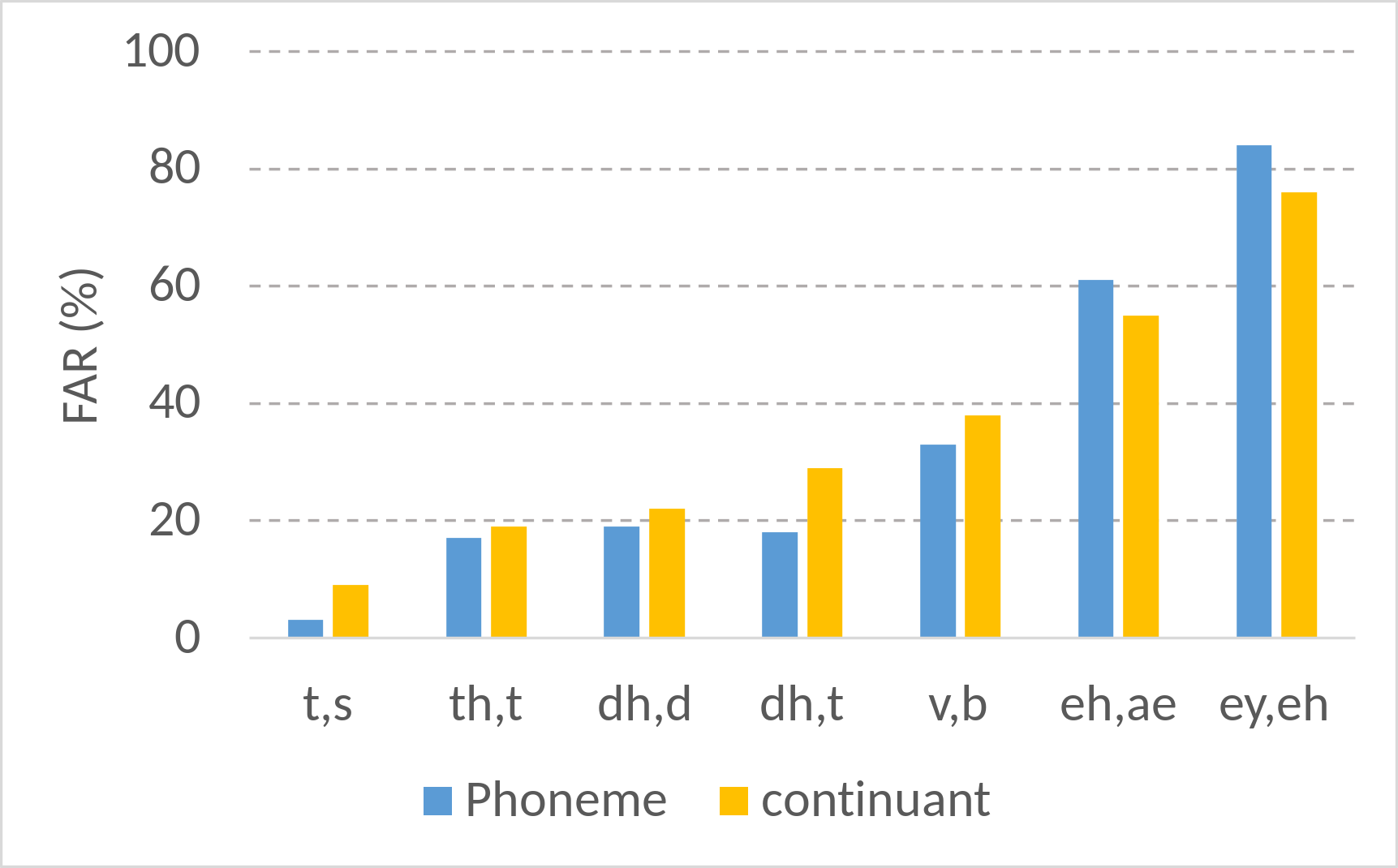}
     \end{subfigure}
     \hfill
     \begin{subfigure}[b]{0.49\textwidth}
         \centering
         \includegraphics[width=\textwidth]{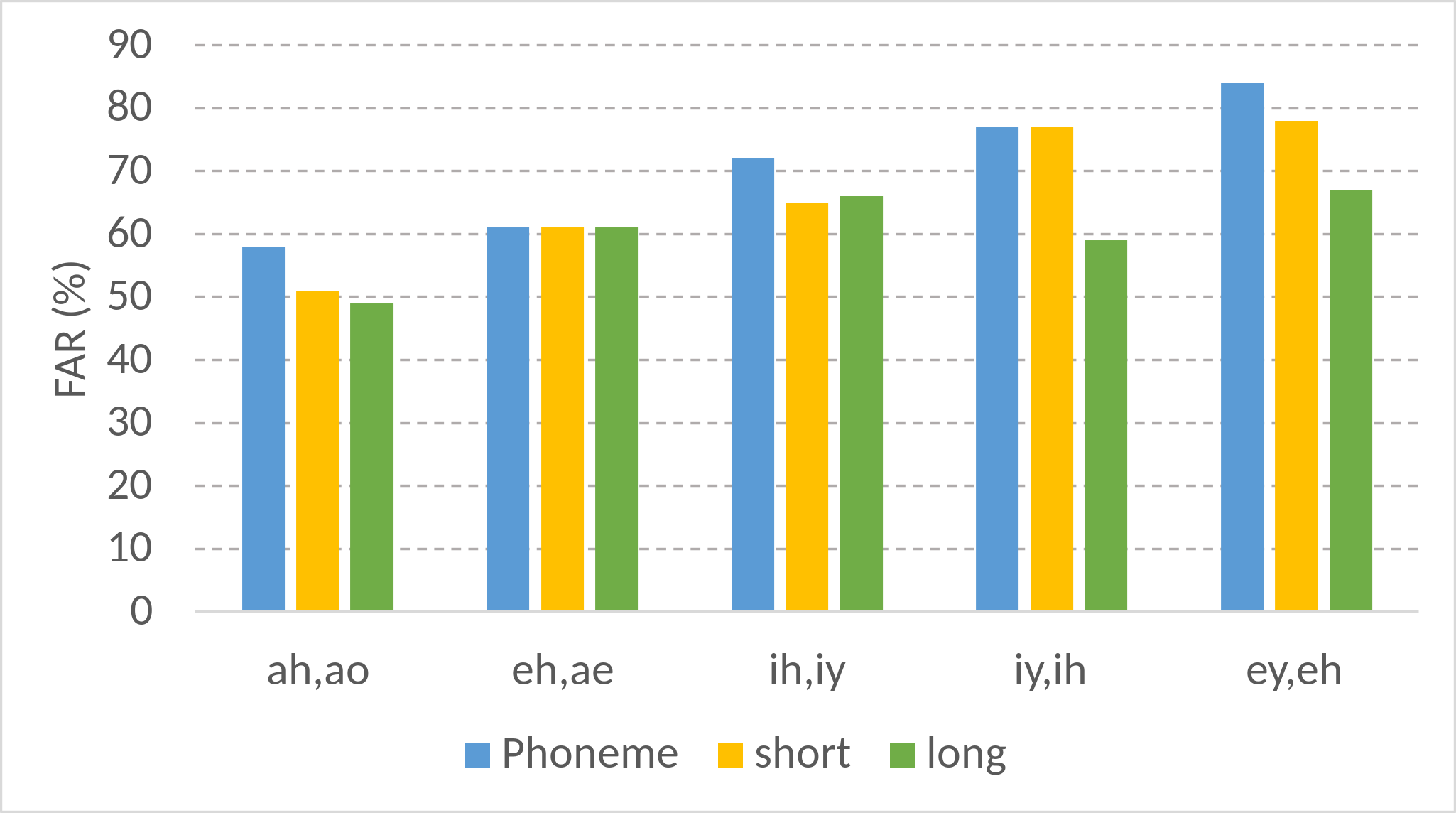}
     \end{subfigure}
        \caption{The FAR of substitution errors computed from phoneme-level assessment and speech attribute-level assessment using attributes that can discriminate between the expected and mistaken phonemes. The voiced attribute improves the discrimination between (/d/,/t/), (/p/,/b/), (/v/,/f/) and (/k/,/g/), while the diphthong improves the discrimination between (/ey/,/eh/), (/ow/,/ao/), and (/aw/,/aa/).}
        \label{fig:res9}
\end{figure}

Figure \ref{fig:res9} shows the FAR of each substitution error computed from phoneme-level assessment and speech attribute-level assessment using attributes that can discriminate between the expected and mistaken phonemes. For instance, the top-left bar graph shows the FAR of pairs of confused phonemes using the phoneme-based model and the voiced attribute as the only attribute that discriminates between each pair.

The results show that the voiced attribute reduced the FAR of “/d/, /t/”, “/p/, /b/”, “/v/, /f/” and “/g/, /k/” compared to the phoneme-based model, however, it failed to effectively discriminate between /jh/ and /ch/ and /z/ and /s/. On the other hand, the dental attribute significantly reduced the FAR of /th/ that was mispronounced as /s/ from ~72\% when using the phoneme-based model to ~37\%. Also, the fricative attribute reduced the FAR of “/v/, /b/” substitution error by 54\%.
The results shown here are promising. They indicate that the speech attributes model, in addition to providing low-level diagnostic information, can also improve phoneme-level mispronunciation detection.
\clearpage

\section{Conclusion}
This paper introduced a novel MDD method to detect the existence of pronunciation errors and provide comprehensive diagnostic information. Unlike the current SOTA phoneme-level MDD which can only recognise the incorrect phoneme, our method, in addition to locating the mistaken phoneme, gives a detailed description of the pronunciation error in terms of which articulators are involved in the error’s production.

This was achieved by first modelling the speech attribute (phonological) features, which include the manners and places of articulation. Given the recent superiority of the pre-trained speech representation model in different downstream tasks, we adopted the wav2vec2 pre-trained model as the core architecture of our speech attribute detection model. The results show that the large wav2vec2 model, which was pre-trained using data from different domains (clean, noisy, telephone, crowd-sourcing), outperformed single-domain models (see Figure \ref{fig:res1}). 

We further proposed a novel multi-label variant of the CTC loss function (SCTC-SB) to handle the non-mutually exclusive nature of speech attributes. This enables the use of a single network for the joint modelling of all speech attributes, making the model efficient in speed and memory. This is in contrast to current methods which require separate models for each attribute \cite{RN148, RN149} or a group of mutually exclusive attributes \cite{RN160,RN161}. Furthermore, the results demonstrate the superiority of the proposed speech attribute detection model over End2End Deep Speech 2-based models \cite{RN137} for the detection of nasality \cite{RN162} and manner of articulations \cite{RN169}. (See Section \ref{sssec:Comp})

The resultant speech attribute detection model was then used in performing MDD on non-native English speech corpus collected from 24 speakers of 6 different native languages.  The experimental results show that our speech attribute-based MDD can achieve a reasonable performance (with $FAR$ below 30\% and $DER$ below 10\%) when trained solely on native speech corpora and applied to non-native speech (see Figure \ref{fig:res6}). On the other hand, phoneme-level MDD which was trained and tested on the same native and non-native datasets achieved $FAR$ and $DER$ of ~57\% and ~31\% respectively (see Table \ref{tab:res1}). This is a significant advantage of the system as one of the major impediments in developing an accurate MDD is the scarcity of the non-native training dataset. 

Furthermore, adding 2 hours of annotated non-native data with 4 hours of native data achieved an average correct detection rate of 88\% and an average diagnosis accuracy of 97\% in the speech attribute MDD. (see Figure \ref{fig:res6})

Moreover, for certain frequently occurring substitution error pairs, employing a single speech attribute yields higher discrimination accuracy compared with phoneme models. For instance, utilizing the voicing attribute proves more effective in distinguishing between the sounds /d/ and /t/ than their phonetic acoustic models.
In addition to a higher accuracy compared to phoneme-based MDD, speech attribute-based MDD provides a low-level description of the pronunciation errors that is directly related to the articulatory system allowing formative feedback to be constructed.

Our future work includes leveraging the universal nature of the phonological features and incorporating speech corpora from multiple languages to train the speech attribute detection model. This will add more variations to the model leading to more robust MDD system. Furthermore, we are planning to extend the MDD system to more challenging domains such as adult and child disordered speech.

\clearpage

\clearpage
\bibliographystyle{elsarticle-num}
\bibliography{refs}
\end{document}